\documentclass[journal]{IEEEtran}
\usepackage{amsmath}
\usepackage{amssymb}
\usepackage{amsthm}
\usepackage{graphicx}
\usepackage{booktabs}
\usepackage{enumitem}
\usepackage{hyperref}

\setlist{nosep, leftmargin=*}

\newtheorem{definition}{Definition}

\begin{document}

\title{Temporal Fair Division in Multi-Agent Systems: From Precise
Alternation Metrics to Scalable Coordination Proxies}

\author{Nikolaos~Al.~Papadopoulos,
        Ismael~T.~Freire,
        Mart\'i~Sanchez-Fibla,
        and~Konstantinos~E.~Psannis
\thanks{N. Al. Papadopoulos (corresponding author, ORCID 0000-0003-1842-8227,
e-mail: nikolaos.papadopoulos@uom.edu.gr) and K. E. Psannis (ORCID
0000-0003-0020-6394) are with the Department of Applied Informatics,
University of Macedonia, Thessaloniki, Greece.}
\thanks{I. T. Freire (ORCID 0000-0002-5740-1513) is with the Institute of
Intelligent Systems and Robotics (ISIR), Sorbonne Universit\'e, Paris, France.}
\thanks{M. Sanchez-Fibla (ORCID 0000-0001-5725-1984) is with the Department of
Information and Communication Technologies, Universitat Pompeu Fabra,
Barcelona, Spain.}}

\maketitle

\begin{abstract}
Many multi-agent games and simulated environments involve several
independent, often learning, agents repeatedly competing for a shared
resource. Conventional fairness measures assess whether resources are
shared equally overall, but cannot distinguish orderly turn-taking from
irregular access patterns that produce long, unpredictable waiting times
despite similar cumulative outcomes. We introduce Rotational Periodicity,
a family of computationally efficient metrics combining two complementary
measurements, the regularity of waiting times between successful accesses
and the balance of access frequencies across agents. We evaluate them
against a more detailed family of alternation metrics in a repeated
threshold-congestion game, the Honey-Jar Game, in which two to ten
reinforcement-learning agents compete for exclusive access to a shared
high-value resource. Our experiments show that independently trained
Q-learning agents often coordinate substantially worse than random-policy
agents, even though conventional fairness metrics consistently report
favourable outcomes. Every member of the family closely reproduces the
rankings of the more computationally expensive alternation metrics, while
computing six to twenty times faster as the number of agents
grows. A direct timing study, which separates the two quantities that drive that
cost, measures the advantage growing in proportion to the number of agents out
to two thousand agents. These findings show that evaluating multi-agent
learning systems
requires temporally aware measures of coordination, not only aggregate
outcomes, and that efficient proxies make this practical for larger
multi-agent games and simulations.
\end{abstract}


\section{Introduction}
\label{sec:intro}

Classical fair division asks \emph{who should receive what}, namely how to partition goods
among agents so that no one envies another's share~\cite{foley67}, each agent
receives a proportional part~\cite{steinhaus48}, or collective welfare is
maximised~\cite{caragiannis19}. A rich literature in economic theory and algorithmic
game theory~\cite{bramsT96,moulin03} has formalised these criteria and established
when and how they can be achieved, almost always treating allocation as a one-shot
problem decided at a single point in time.

Yet many environments of practical significance are \emph{repeated}: agents
compete for the same limited resource episode after episode, and what
matters is whether every agent obtains it \emph{often enough and regularly
enough} over the long run, beyond who obtains it on any given day.
Consider $n$ autonomous vehicles sharing a single-lane bridge, where a
controller that grants passage fairly \emph{on average} but allows one
vehicle to monopolise access for long stretches creates unacceptable
waiting times even with balanced cumulative counts. Network bandwidth
allocation, smart-grid energy distribution, traffic signal control, and
turn-based collaborative robotics all share this
structure~\cite{izmirlioglu24,mota24}, recurring wherever game-playing
agents are distributed across autonomous, learning components, from
non-player-character swarms coordinating shared virtual space to game
servers scheduling a shared channel, making temporal coordination a
game-design concern in its own right. The relevant fairness criterion here
is \emph{temporal}: an allocation scheme is temporally fair if the sequence
of wins for each agent, as a whole, satisfies some notion of balance and
regularity. This dimension has received growing attention under several
labels, online fair division~\cite{he19}, repeated
allocation~\cite{bouveret11}, sequential resource sharing~\cite{perolat17,leibo17},
yet quantifying \emph{how close} an observed sequence is to the temporal
fairness ideal has received comparatively little attention. How do we
measure this proximity in a way sensitive to coordination failures and
tractable for large populations?

The present paper addresses these questions by studying temporal fair division in
the Honey-Jar Game (HJG; the Multi-agent Battle of the Exes, MBoE, in an earlier conference paper), a repeated competitive game in which
$n$ self-interested agents repeatedly attempt to claim a shared high-reward resource.
We make three contributions.

\emph{Contribution 1: Framework.}
We formalise HJG as a repeated fair division instance and identify \emph{Perfect
Alternation} (PA), a Pareto-optimal coordination regime in which every agent wins
exactly once in every sliding window of $n$ consecutive episodes, as the canonical temporally fair solution
(a Nash equilibrium in the two-agent ballistic formulation~\cite{papadopoulos26},
a single simultaneous move/stay decision per episode, rather than the
multi-round race studied experimentally here).
We show that PA satisfies temporal proportionality and a form of temporal
envy-freeness analogous to the classic EF notion in static fair division.

\emph{Contribution 2: Rotational Periodicity.}
We formalise and empirically validate Rotational Periodicity
(RP)~\cite{papadopoulos25rp}, a family of lightweight metrics
that decompose temporal fairness into two orthogonal dimensions: a \emph{rhythm}
component (RS), measuring how regularly each agent's inter-win gaps match the ideal
$n-1$ episodes, and a \emph{frequency} component (WPE), measuring whether each
agent's win count approximates the ideal $\nu/n$. All variants are normalised against
the PA baseline and computed in $O(\nu + n)$ time, making RP the practical choice for
large populations where ALT becomes intractable. Full definitions and extensions
(weighted FRP, equitable ERP) are given in Section~\ref{sec:rp}.

\emph{Contribution 3: Comparative Empirical Study.}
We present a systematic comparison of RP with the ALT family of
sliding-window metrics~\cite{papadopoulos21,papadopoulos26} across agent populations
$n \in \{2, 3, 5, 8, 10\}$, using both Q-learning policies and random baselines.
The comparison spans metric expressiveness, sensitivity, and computational cost,
and leads to concrete guidance on when each family is preferable. A direct
timing study carries the cost comparison far past the populations a learning
experiment can reach, separating the effect of the agent count from that of the
episode budget and measuring the advantage out to $n=2000$.

Our experiments reveal a consistent pattern. For $n \geq 3$, Q-learning agents
achieve high Reward Fairness ($>0.92$) yet score \emph{worse than random
policies} on RP by $3$--$120\%$ and on CALT by $7$--$35\%$ (across all four
state/reward configurations; the Type-A/ILF condition of
Table~\ref{tab:coordination_scores} alone gives $21$--$120\%$/$11$--$35\%$),
both gaps peaking at $n=3$. At $n=2$ the agents converge to a monopoly,
a distinct regime traditional metrics do detect. This \emph{coordination
gap}, invisible to traditional metrics elsewhere, motivates temporal
fairness measures alongside standard ones. The claim is about measurement
rather than about the limits of multi-agent learning. The learned policies
are the instrument that exposes what aggregate metrics miss, and a stronger
algorithm would give a different reading, not a different verdict on the
metrics.

\subsection{Related Work}
\label{sec:related}

\subsubsection{Static Fair Division}

The theoretical study of fair allocation traces back to the ``problem of fair division''
posed by Steinhaus~\cite{steinhaus48} and formalised through the notions of
proportionality and envy-freeness~\cite{foley67}. For divisible goods (``cake-cutting'')
a rich literature establishes existence and complexity results~\cite{bramsT96}.
For indivisible goods (closer to our setting), Lipton et al.~\cite{lipton04} introduced
Envy-Freeness up to one Good (EF1), showing it is always achievable, while
subsequent work established stronger notions and their computational properties~\cite{caragiannis19}.
The textbook by Moulin~\cite{moulin03} provides a comprehensive treatment of fair
division from a welfare economics perspective.

\subsubsection{Sequential and Repeated Allocation}

When items must be allocated one at a time over multiple rounds, round-robin
mechanisms are natural candidates. Bouveret and Lang~\cite{bouveret11} show
that $n$-periodic round-robin achieves approximate proportionality for
indivisible goods. Adams and Segal-Halevi~\cite{adams2026} seek balanced
sequences of permutations in which each agent receives each item exactly
once over an $n$-round cycle (a Latin-square structure), directly analogous
to our Perfect Alternation benchmark. He et al.~\cite{he19} investigate
fairness in an online setting where items arrive sequentially and
allocations must be made irrevocably; recent work extends this to requiring
fairness at every round prefix, not only at the end~\cite{elkind2025,igarashi2024}.
Since exact envy-freeness is rarely attainable with indivisible items, this
literature works with bounded relaxations, EF1, its ordinal strengthening
SD-EF1, and proportional up to one good (PROP1). Cookson et
al.~\cite{cookson2025} prove an allocation that is SD-EF1 per day and PROP1
overall always exists; Choi and Li~\cite{choi2026} add scheduling, showing a
buffer of $n/2$ rounds suffices for temporal EF1 under identical days while
stronger notions (TEFX, TMMS) remain impossible even then. These works
establish theoretical limits on guaranteed temporal fairness; our
contribution is complementary and empirical, quantifying how closely an
observed sequence approaches the ideal. Our setting also differs in that the
same resource is competed for every round, so allocation is genuinely repeated rather
than sequential.

\subsubsection{Turn-Taking in Multi-Agent Systems}

Turn-taking as an emergent coordination behaviour has been studied in both
biological and artificial agent settings~\cite{rankin07,dejong08}. The Battle of
the Exes (BoE), introduced by Hawkins and Goldstone~\cite{hawkins16}, provides
an idealised game in which two agents must learn to alternate access to a
high-reward location without communication. Prior work~\cite{papadopoulos21} extended this game to $n$ agents (HJG)
and proposed the ALT family of metrics for evaluating turn-taking quality.
Raffensperger et al.~\cite{raffensperger11} proposed a simpler metric for
emergent turn-taking in communication experiments.
Freire et al.~\cite{freire2020,freire2023} study convention formation and theory-of-mind
models in dyadic coordination games; Gasparrini and Sanchez-Fibla~\cite{gasparrini2018}
show that loss aversion promotes turn-taking among independent Q-learners in BoE-like
settings; and Puig Camps~\cite{puig2018} confirms that turn-taking is a typical
emergent outcome in computational BoE variants.

HJG's essential structure, a single resource whose value collapses under overuse,
situates it within a broader family of congestion and anti-coordination games.
Rosenthal~\cite{rosenthal73} introduced congestion games, in which each player's
payoff depends on how many others share the same resource, and showed that a
potential function guarantees the existence of pure-strategy Nash equilibria.
Market-entry games~\cite{selten82} study a closely related setting in which $n$
players simultaneously decide whether to enter a market of limited capacity, with
payoffs collapsing once the number of entrants exceeds it; experimental work in
this tradition~\cite{rapoport98} finds that aggregate entry converges near the
capacity-efficient level even though individual behaviour is difficult to predict
in advance. The El Farol Bar Problem~\cite{arthur94} and its formal abstraction,
the Minority Game~\cite{challet97}, capture the same anti-coordination tension for
boundedly rational, inductively reasoning agents repeatedly choosing between two
options. These formulations are largely stateless and simultaneous-move, and
they ask how many agents enter. HJG instead embeds the contest in a minimally
dynamic environment in which agents' spatial approach toward the resource is
observable before commitment, a structural choice made so that movement
itself can function as an implicit coordination signal; Goldstone
and colleagues~\cite{goldstone04} show that such movement indeed allows
groups to self-organise access to a shared resource without explicit
communication. This structural choice is separate from a further,
evaluation-level gap that holds regardless of which reward rule governs
collisions: even where this literature's own constructions repeat across
rounds, as the El Farol Bar Problem and the Minority Game do, they track an
aggregate statistic, not a specific agent's identity over time, while our
evaluation asks directly which agent obtains access and when across a
repeated sequence of contests.

RP was first proposed as a scalable complement to ALT in a conference
paper~\cite{papadopoulos25rp}. A related study~\cite{papadopoulos26}
conducts a large-scale study of coordination
failure in HJG, showing that Q-learning policies consistently underperform
random baselines on ALT metrics across all tested configurations. The present
work builds on both. It extends the conference proposal of RP with a full
formalisation and a fair division framing, and complements that study with a systematic comparison of RP against the full
ALT family across $n \in \{2,3,5,8,10\}$.

\subsubsection{Multi-Agent Reinforcement Learning and Common-Pool Resources}

Multi-agent reinforcement learning (MARL) has been applied extensively to resource
sharing and common-pool problems~\cite{perolat17,leibo17}, where a persistent
challenge is that standard reward signals do not incentivise coordination
beyond cumulative reward: Perolat et al.~\cite{perolat17} show that agents
in common-pool resource games learn behaviours that deplete the resource
despite high individual returns, echoed here by High Reward Fairness
coexisting with coordination failure. A related tension concerns the
computational cost of tracking fairness over time: Kumar and
Yeoh~\cite{kumar2025} show enforcing history-dependent (perfect-recall)
fairness inflates the state space unboundedly with the horizon, proposing
past-discounting (geometric decay over older allocations) to keep learning
tractable, while RP addresses the analogous measurement-side tractability
problem by summarising each agent's history through inter-win gaps alone in
$O(\nu+n)$ time.
Two recent papers in this journal address closely related coordination
mechanisms in MARL. Yu et al.~\cite{yu2026} learn an explicit coordination-graph
structure so agents make dependency-aware joint decisions instead of acting
independently, and Yang et al.~\cite{yang2025} learn individual potential-based
rewards so that agents receiving only a shared team reward still develop
diversified individual policies. Both speak to the same individual-versus-aggregate
tension our own findings expose, that a healthy-looking aggregate outcome can
mask individually uncoordinated behaviour.

\section{Materials and Methods}
\label{sec:methods}

\subsection{Experimental Design}
\label{sec:experimental-design}

\subsubsection{The Honey-Jar Game as Temporal Fair Division}
\label{sec:problem}

\emph{Game Formulation.}

The game was introduced in an earlier conference paper~\cite{papadopoulos25rp} as
the Multi-agent Battle of the Exes (MBoE), one possible, non-exclusive reading of
its payoff structure, inspired by BoE~\cite{hawkins16} rather than a strict
multi-agent generalisation of it. Starting from BoE's anti-coordination and
alternation objectives, the minimally dynamic implementation with graded collision
payoffs makes the resulting game, in essence, a congestion game; we therefore refer
to it as the \emph{Honey-Jar Game} (HJG) throughout the remainder of this paper.
(Incidentally, the acronym MBoE is preferably read as \emph{Multi-agent Benefit
of Exclusivity}.)

The Honey-Jar Game with $n$ agents proceeds in discrete
episodes $t = 1, 2, \ldots, \nu$. At each episode $t$, every agent $i \in \{1,\ldots,n\}$
independently attempts to reach a terminal position. The payoff structure is:
\begin{itemize}
  \item A \emph{solo winner} (the unique agent reaching its terminal position) receives $r_\mathrm{high}$.
  \item In a \emph{partial tie} ($2 \leq m < n$ agents arriving simultaneously), each tied
    agent receives a fixed fractional share, $r_\mathrm{high}/n$ under \emph{Inverse Linear
    Fractional} (ILF) rewards, or $r_\mathrm{high}/n^2$ under \emph{Inverse Quadratic
    Fractional} (IQF) rewards.
  \item If \emph{all} $n$ agents reach their terminal positions simultaneously, every agent
    receives $0$, since full congestion destroys the resource.
  \item Agents that do not reach their terminal position receive zero.
\end{itemize}
Payoffs are therefore non-increasing in the number of simultaneous claimants
($r_\mathrm{high}$, then a fixed fractional share, then $0$), a
threshold-congestion structure with a capacity collapse at full load. Under ILF
the tied share $r_\mathrm{high}/n$ is exactly the per-capita split of the
resource across the whole population; IQF adds a further congestion penalty.

The game admits a simple picture, giving it its name: a group of bear cubs,
unable to communicate, share a single jar of honey, each stood a few steps
away so that approaching or holding back is visible to the others, and
movement itself acts as a commitment signal; the only collectively optimal
behaviour is spontaneous turn-taking. We set $r_\mathrm{high} = 100$
throughout. Agents observe the game state and update a Q-table according to
the standard tabular Q-learning rule with learning rate $\alpha = 0.3$,
discount factor $\gamma = 0.999$, and an $\varepsilon$-greedy policy with
$\varepsilon$ decaying linearly from 0.9 to 0.004 over $75\%$ of the episode
budget.

We use a tabular representation deliberately. With values stored explicitly,
the formation or absence of a turn-taking policy can be read directly off the
Q-table as it is built, and a coordination failure attributed to the learning
dynamics rather than to a function approximator's representational limits.
Since the object of study here is how coordination is \emph{measured}, an
inspectable learner is worth more than a stronger opaque one. Two further
considerations fix the choice. Independent Q-learning is the algorithm under
which the ALT family was developed and previously measured on this
game~\cite{papadopoulos21,papadopoulos26}, so holding it constant makes this a
comparison of metrics rather than a confound between metric and learner. It is
also a documented positive case: independent learners are known to coordinate
poorly in repeated games~\cite{shoham08}, and showing that aggregate metrics
conceal a coordination failure requires a setting in which one is present.
Whether the metrics are sound does not rest on that, which is what
Section~\ref{sec:criterion-validity} establishes independently of any learner.

\emph{Connection to fair division.}
At each episode the right to be the sole winner (the ``high-value resource'') can
be allocated to at most one agent. Over $\nu$ episodes the resource is available
$\nu$ times; perfect efficiency requires that it is claimed every episode, while
perfect fairness requires that each agent claims it the same number of times.
This is precisely a repeated fair division problem in which the item of value is
the exclusive access right, and the allocation is determined endogenously by the
agents' strategies.

\emph{Two State Representations.}

\textbf{Type-A} states encode only agent positions: $s_t = [p_1^t, \ldots, p_n^t]$.
\textbf{Type-B} states additionally include a memory vector recording, for each
agent $i$, whether it was the sole winner in the previous episode:
$s_t = [p_1^t, \ldots, p_n^t, z_1^t, \ldots, z_n^t]$, where $z_i^t \in \{0,1\}$.
Type-A serves as the primary experimental condition; Type-B is included for
completeness.

\emph{Perfect Alternation as the Temporal Fairness Ideal.}

\begin{definition}[Perfect Alternation Equilibrium]
An outcome is said to satisfy \emph{Perfect Alternation} (PA) if, in every
\emph{sliding} window of $n$ consecutive episodes, each agent is the sole winner
exactly once. Equivalently, the sequence of winners forms an $n$-periodic cycle in
which every agent appears exactly once per period.
\end{definition}

PA is Pareto-optimal and constitutes a Nash equilibrium in the two-agent
ballistic formulation~\cite{papadopoulos26} (a single simultaneous move/stay
decision per episode, distinct from the multi-round race that HJG's
experiments actually study); for $n > 2$ it serves as the
canonical temporally fair reference point. PA satisfies the following temporal
analogues of classical fair division criteria:

\textit{Temporal Proportionality.} Each agent wins $\nu/n$ times in $\nu$
episodes, receiving exactly $1/n$ of the total high-value resource.
\textit{Temporal Envy-Freeness.} Since all agents win at the same rate with
the same inter-win gaps, no agent $i$ strictly prefers the win-history of
any other agent $j$ to its own, so no temporal envy arises.
\textit{Regularity.} The inter-win gaps for each agent are identically equal
to $n-1$, minimising uncertainty about resource access timing.

These properties hold under the symmetric single-resource valuation of HJG, where
every agent values the contested resource equally, the sole winner of an episode
receives $r_\mathrm{high}$, and temporal envy is assessed over cumulative wins at
$n$-episode cycle boundaries. Under this valuation PA is not a single outcome but an
\emph{equivalence class}. The $n!$ periodic winner-sequences (one per permutation of
a block) all satisfy the definition, and every metric assigns them the maximal score,
since they share the same per-agent inter-win gaps ($n-1$) and win counts ($\nu/n$).
PA is \emph{sufficient} for temporal proportionality and for envy-freeness at cycle
boundaries, but not \emph{necessary}. Outcomes that equalise wins per block without a
fixed period (for $n=2$, the clumped sequence $A,B,B,A,\ldots$) are also proportional
and cycle-boundary envy-free. What additionally characterises PA is \emph{regularity},
namely constant inter-win gaps equal to $n-1$; PA is thus the maximally regular member
of the temporally proportional class. A full axiomatic characterisation, and the
heterogeneous-valuation case, are left to future work and partially addressed through
the weighted RS and ERP variants introduced below.

The PA concept is analogous to $n$-periodic round-robin allocation studied in the
sequential allocation literature~\cite{bouveret11}, where each agent takes exactly one turn
per round of $n$ slots. The key difference is that in HJG, the alternating
schedule must \emph{emerge} from independent agent learning rather than being
enforced by a central planner.

\subsubsection{Metrics for Temporal Fairness}
\label{sec:metrics}

We present the full set of metrics in the framework, progressing from the coarsest
to the finest-grained.

\emph{Traditional Metrics.}

Let $R_i^\nu$ denote agent $i$'s cumulative reward through episode $\nu$, and let
$R_\mathrm{max} = \max_i R_i^\nu$.

\begin{definition}[Efficiency]
\[
  E = \frac{\sum_{i=1}^n R_i^\nu}{\nu \cdot r_\mathrm{high}}.
\]
$E$ reaches $1.0$ when no episode is wasted (solo winner every episode,
no ties with shared rewards). Under PA, $E = 1.0$.
\end{definition}

\begin{definition}[Reward Fairness]
\[
  \mathrm{RF} = \frac{\sum_{i=1}^n R_i^\nu}{n \cdot R_\mathrm{max}}.
\]
$\mathrm{RF}$ approaches $1.0$ when rewards are evenly distributed. Under PA,
$\mathrm{RF} = 1.0$. As we demonstrate empirically, high RF is a necessary but
far from sufficient condition for temporal fairness. Random policies can achieve
RF~$> 0.9$ without any coordination whatsoever.
\end{definition}

\emph{ALT Metrics: A Sliding-Window Family.}

The ALT family~\cite{papadopoulos21,papadopoulos26} evaluates turn-taking quality
by sliding a window of width $n$ across the episode sequence and averaging a
batch score $\beta_j$ over all $\nu - n + 1$ windows. Formally:
\[
  \mathrm{ALT} = \frac{1}{\nu - n + 1} \sum_{j=0}^{\nu - n} \beta_j.
\]
The six variants differ in their definition of $\beta_j$.
Let batch $j$ span episodes $[j, j+n-1]$. Define:
$f_j$: number of distinct agents reaching their terminal position at least once in batch $j$;
$t_j$: total terminal arrivals in batch $j$;
$w_j$: number of episodes with exactly one winner in batch $j$;
$g_j$: number of agents achieving exactly one \emph{solo} win in batch $j$;
$y_k$: number of agents reaching their terminal position in episode $k$ of batch $j$.

\begin{table}[t]
\caption{ALT Metric Variants and Their Batch Score Formulas}
\label{tab:alt_variants}
\small
\centering
\renewcommand{\arraystretch}{1.6}
\setlength{\tabcolsep}{3pt}
\begin{tabular}{cp{2.5cm}p{4.2cm}}
\toprule
\textbf{Metric} & \textbf{$\beta_j$ formula} & \textbf{Interpretation} \\
\midrule
\multicolumn{3}{l}{\textit{Primary metrics (CALT, EALT, AALT)}} \\[3pt]
CALT  & $\dfrac{\beta_j^{\mathrm{qFALT}}}{n(n-1)} \,\times$ \newline $\sum_{k=1}^{n}(n - y_k)$ & Tie penalty weighted by reach diversity ($\beta_j^{\mathrm{qFALT}}$, defined below); most comprehensive single-number summary \\[4pt]
EALT  & $\dfrac{w_j \cdot f_j}{n^2}$   & Exclusive-win episodes weighted by reach diversity \\[4pt]
AALT  & $g_j / t_j$ & Fraction of agents achieving exactly one solo win per window \\
\midrule
\multicolumn{3}{l}{\textit{Auxiliary variants (additional nuance near PA)}} \\[3pt]
FALT  & $f_j / t_j$ & Fraction of terminal arrivals accruing to distinct agents \\
qFALT & $(f_j/t_j)^2$ & FALT ($\beta_j^{\mathrm{FALT}}$) squared, amplifying sensitivity near PA \\
qEALT & $(\beta_j^{\mathrm{EALT}})^2$ & EALT (above) squared, amplifying sensitivity near PA \\
\bottomrule
\end{tabular}
\renewcommand{\arraystretch}{1.0}
\end{table}

Every ALT metric equals $1.0$ under PA and $0$ under complete coordination failure.
The three primary metrics (CALT, EALT, AALT) capture complementary aspects of
coordination quality at different strictness levels: CALT applies the per-episode
tie penalty $\sum_k(n-y_k)$ weighted by the reach-diversity score
$\beta_j^{\mathrm{qFALT}}$, making it the most comprehensive single-number summary;
EALT rewards exclusive-win episodes weighted by how many distinct agents reached
the terminal at all; AALT measures per-window agent coverage. The auxiliary
variants (FALT, qFALT, qEALT) provide additional discrimination near-perfect
alternation and are not the focus of this paper. All ALT computation has
complexity $O(\nu \cdot n)$.

\textbf{AltRatio and PA-Equivalent Agents.}
Following~\cite{papadopoulos26}, the \emph{AltRatio} is $x/n$, the proportion of
agents behaving as if in perfect alternation. For CALT specifically, the quadratic
construction of its batch score means the metric scales approximately as $(x/n)^2$
(with a near-zero intercept $\varepsilon \approx 0$~\cite{papadopoulos26}), so:
\[
  \mathrm{AltRatio}_\mathrm{CALT} \approx \sqrt{\mathrm{CALT}},
\]
\[
  \text{\emph{PA-equivalent agents}} \approx n\sqrt{\mathrm{CALT}}.
\]
For EALT the scoring is linear in $x/n$, so $\mathrm{AltRatio}_\mathrm{EALT} =
\mathrm{EALT}$ directly (no square-root transformation). For AALT the
relationship is piecewise linear with a threshold at $\mathrm{AltRatio} =
0.5$~\cite{papadopoulos26}, so $\mathrm{AltRatio}_\mathrm{AALT} = \mathrm{AALT}$
remains a direct reading away from that threshold, without a square-root
transformation.

\emph{Rotational Periodicity: A Linear-Time Family.}
\label{sec:rp}

RP metrics measure temporal fairness by analysing each agent's \emph{waiting pattern}
individually (specifically, the sequence of gap lengths between consecutive wins) rather than
by scanning the global episode sequence in batches.

\textbf{Waiting Periods.}
For agent $i$, let $w_{i,1} < w_{i,2} < \cdots < w_{i,k_i}$ denote the episodes
of its $k_i$ solo wins over a run of $\nu$ episodes (numbered $0,\ldots,\nu-1$).
If $k_i \geq 1$, the agent's \emph{waiting periods} are the $k_i - 1$ inter-win
gaps $w_{i,j+1} - w_{i,j} - 1$ (episodes strictly between two consecutive
wins) for $j = 1,\ldots,k_i-1$, together with a \emph{leading} period of
length $w_{i,1}$ before its first win (included only if positive) and a
\emph{trailing} period of length $\nu - 1 - w_{i,k_i}$ after its last win
(included only if positive). If $k_i = 0$, the entire run counts as a single
waiting period of length $\nu$. Counting the leading and trailing periods
reflects the original waiting-period design~\cite{papadopoulos25rp}. Time
spent before an agent's first win or after its last win is genuine waiting,
not a boundary artefact to be discarded. Let $\bar{r}_i$ denote the mean
length of these waiting periods.

\textbf{Rotational Score (RS).}
Let $r_i^* = n - 1$ be the ideal waiting-period length under Perfect
Alternation. The \emph{Rotational Score} is
\[
  \mathrm{RS}_i = \frac{\min(\bar{r}_i,\, r_i^*)}{\max(\bar{r}_i,\, r_i^*)}.
\]
RS replaces the asymmetric AWE sub-measure used in an earlier conference
paper~\cite{papadopoulos25rp}, which applied a hard threshold at
$\bar{r}_i = 2r_i^*$ (returning $0$ for any gap beyond twice the ideal)
and gave the same penalty to an agent waiting $2r_i^*$ as to one waiting
$100\,r_i^*$. RS penalises deviations \emph{symmetrically}. An agent
winning at twice the ideal rate ($\bar{r}_i = r_i^*/2$) and an agent
winning at half the ideal rate ($\bar{r}_i = 2r_i^*$) both receive
$\mathrm{RS}_i = 0.5$. Under PA, $\bar{r}_i = r_i^* = n-1$ and
$\mathrm{RS}_i = 1$. As $\bar{r}_i$ diverges from $r_i^*$ in either
direction $\mathrm{RS}_i$ decreases continuously towards $0$, providing a
non-trivial signal even for very infrequent winners.

\textbf{Weighted Temporal Fair Division.}
In heterogeneous-priority settings where agent $i$ holds target share $w_i > 0$
($\sum_{i=1}^n w_i = 1$), the ideal gap generalises to
$r_i^* = 1/w_i - 1$ (the expected inter-win gap when the agent wins $w_i$
of all episodes). The resulting \emph{weighted RS},
$\mathrm{RS}_i^w = \min(\bar{r}_i, r_i^*)/\max(\bar{r}_i, r_i^*)$,
equals $1$ when the agent's empirical win rate matches its priority $w_i$
exactly. The uniform case ($w_i = 1/n$, $r_i^* = n-1$) is recovered as
a special case. This formulation aligns RS directly with \emph{weighted
proportionality}, the standard fairness criterion for agents with
heterogeneous claims~\cite{moulin03}. Each agent's share of the contested
resource should equal $w_i$.

\textbf{Waiting Periods Evaluation (WPE).}
WPE measures whether each agent's \emph{frequency} of wins is consistent with
PA. Let $t_i$ denote the number of win events recorded by agent $i$; under PA
every agent wins exactly $t_i^* = \nu/n$ times. Then:
\[
  \mathrm{WPE}_i =
  \begin{cases}
    1 - \dfrac{|t_i - t_i^*|}{t_i^*} & \text{if } t_i < 2t_i^*, \\
    0 & \text{otherwise.}
  \end{cases}
\]
For an under-winning agent ($t_i \leq t_i^*$) the score reduces to $t_i/t_i^*$,
the fraction of its fair share of wins actually obtained. WPE thus captures the
\emph{distribution} of access opportunities, namely whether the agent receives the fair
share of turns in terms of frequency.

Both targets ($r_i^* = n-1$ for RS, $t_i^* = \nu/n$ for WPE) are read directly
off the Perfect Alternation definition above rather than imposed by an
external schedule. HJG has no round-robin-style central planner assigning
turn order; RS and WPE instead measure how closely each agent's
independently learned behaviour approaches these PA-derived targets, giving
the metrics a well-defined notion of distance from the ideal that a
scheduling rule by itself does not supply.

\textbf{Combined RP.}
RP combines exactly two components, not three. RS is not an independent
third quantity. It is the raw waiting-period statistic $\bar r_i$ of the
Waiting Periods paragraph above, normalised into a bounded $[0,1]$ score.
Together, RS and WPE span the two properties that jointly define Perfect
Alternation for an agent: constant inter-win gaps equal to $n-1$ (rhythm),
and exactly $\nu/n$ wins (frequency); an agent satisfies both simultaneously
if and only if it is perfectly alternating. This pairing is fixed by that
definitional requirement, inherited directly from the AWE+WPE
design~\cite{papadopoulos25rp} (RS replacing AWE only in its
scoring rule, above, not in what it measures), and is
not selected by searching over alternative weightings or sub-metric choices
for whichever correlates best with the ALT family. The raw $\bar r_i$
statistic itself is reported separately, alongside RS and WPE, in
Table~\ref{tab:correlations} purely as a diagnostic confirming that the
rhythm signal survives even before normalisation; it is not a candidate third
term, since including both the raw statistic and its normalised score would
double-count the same dimension.

The per-agent RP is a weighted combination:
\[
  \mathrm{RP}_i = \frac{\alpha \cdot \mathrm{RS}_i + \beta \cdot \mathrm{WPE}_i}{\alpha + \beta},
\]
and the system-wide metric is $\overline{\mathrm{RP}} = (1/n)\sum_i \mathrm{RP}_i$.
In all experiments we set $\alpha = \beta = 1$, giving equal weight to rhythm and
frequency; this is not an arbitrary choice among many untested alternatives, as the
paragraph below reports a direct sweep of the weighting against the ALT family
showing the equal-weight point sits inside a wide plateau of equally good choices.
Domain-specific settings can emphasise one component; for example,
time-sensitive applications may prefer $\alpha > \beta$. Checking
$\overline{\mathrm{RP}}$ against CALT, EALT, and AALT (Table~\ref{tab:correlations})
validates this fixed, definitionally-motivated combination against an
independently-established, precise reference family; it does not select the
combination, and the definition above is unaffected by whichever pairing in
that table happens to correlate most strongly in this particular sample.
This distinction matters because the two exercises carry different risks:
using Q-learning's own performance to justify a metric that is then used to
judge Q-learning would be circular (the concern independently raised in the
related study~\cite{papadopoulos26}, Section~\ref{sec:criterion-validity}),
whereas checking a fixed, cheap proxy against an already-validated, expensive
reference measure is standard practice and not circular; the risk there is
instead a milder statistical one, overfitting a small sample by searching
over many candidate combinations, which is why we fix RS+WPE by definition
rather than by search.

\textbf{Robustness, and why RS remains necessary despite it.}
Sweeping the WPE weight $w = \beta/(\alpha+\beta)$ against CALT, EALT, and AALT
on the $N=30$ configurations of Table~\ref{tab:correlations} produces a
plateau. Every $w \in [0.35, 1.0]$ yields the identical mean Spearman
correlation ($0.978$), since Spearman correlation depends only on rank and
RS and WPE never cross in a way that reorders these particular $30$
configurations. The equal-weight choice ($w=0.5$) sits inside this plateau,
so it sacrifices nothing relative to any alternative up to and including
pure WPE ($w=1$). This should not be read as evidence that RS is redundant.
The plateau is a property of this dataset's dominant failure mode, in which
agents that go permanently idle (Section~\ref{sec:criterion-validity}) lose
rhythm and frequency together, so the two sub-measures cannot disagree enough
to change the ranking. Where they can disagree, RS remains essential. In the
$n=2$ worked example below (Section~\ref{sec:worked}), $ABABABAB$ and $ABBAABBA$ share an identical WPE
of $1.0$ (four wins each); WPE alone cannot tell them apart, and only RS
distinguishes clumped from steady access. The synthetic experiments of
Section~\ref{sec:criterion-validity} show the same divergence, in opposite
directions depending on the failure mode: the stuck-agent sweep gives
$\mathrm{RS} > \mathrm{WPE}$ ($0.161$ vs.\ $0.059$ at $k=3$, $\theta=0.9$),
while the ballistic share-imbalance sweep gives $\mathrm{RS} < \mathrm{WPE}$
($0.111$ vs.\ $0.200$ at $w_A=0.9$). RS and WPE therefore capture genuinely
different information in general; they simply happen not to need to disagree
in order to rank the particular Q-learning runs studied here.

\textbf{Win-Event Definition.}
Both WPE and the AWE sub-formula of an earlier conference
paper~\cite{papadopoulos25rp} require specifying what counts as a
\emph{win event} for agent $i$: either (i)~\emph{exclusive wins} (sole
victories, episodes in which exactly one agent reaches its terminal
position), or (ii)~\emph{all terminal reaches} (including simultaneous
arrivals). The two definitions produce systematically different behaviour.
Exclusive wins are sparse for $n \geq 3$. Inter-win gaps routinely exceed
$2r_i^*$, collapsing the AWE formula to $0$ for all Q-learning agents in
our experiments (RS is immune to this collapse because its symmetric ratio
$\min/\max$ has no hard cutoff). All-reaches events, by contrast, occur so
frequently that $t_i \gg 2t_i^*$, collapsing $\mathrm{WPE}_i = 0$ for
$n \geq 3$. The choice therefore depends on what the researcher wishes
to quantify: sole-victory frequency (exclusive wins, measuring individual
dominance) or arrival rhythm relative to all interactions (all reaches,
appropriate when ties carry information). Table~\ref{tab:correlations}
reports Spearman correlations for all six sub-metric variants against the
three ALT primary metrics; WPE with exclusive wins consistently achieves
the strongest alignment ($\rho_S = 0.967$--$0.996$), and the combined
$\overline{\mathrm{RP}}$ is reported for the three most informative pairings.

All four named excl/reach pairings of $\overline{\mathrm{RP}}$ achieve
$\rho_S \geq 0.95$ against every ALT primary metric (Supplementary
Table~S2 lists each pairing individually); we report $\overline{\mathrm{RP}}_\text{excl}$
throughout for concreteness, not because it scores highest.

\textbf{Fair and Equitable Variants.}
\emph{Fair RP} (FRP) uses uniform ideal values $r_i^* = n-1$ and $t_i^* = \nu/n$
for all agents, appropriate when equal treatment is desired. \emph{Equitable RP}
(ERP) uses the weighted variants $\mathrm{RS}_i^w$ and a priority-adjusted WPE
with $t_i^* = w_i \cdot \nu$, enabling deliberate asymmetric allocations
directly analogous to weighted proportional fair division in the classical sense.

\textbf{Computational complexity.}
Computing $\overline{\mathrm{RP}}$ requires a single forward pass over the episode
sequence to identify each agent's win episodes ($O(\nu)$), followed by
per-agent gap calculations ($O(n)$ additional steps). The total complexity is
$O(\nu + n)$, a strict improvement over ALT's $O(\nu \cdot n)$.

\emph{Coordination Score.}

To assess whether a policy achieves coordination \emph{above chance}, we define
a normalised coordination score for any metric $M$:
\[
  \mathrm{CS}(M) = \frac{M_\mathrm{QL} - M_\mathrm{rand}}{1 - M_\mathrm{rand}},
\]
where $M_\mathrm{QL}$ is the metric value under Q-learning and $M_\mathrm{rand}$
under a purely random policy~\cite{papadopoulos26}. A positive CS indicates
above-chance coordination; a negative CS indicates \emph{worse} coordination
than a random baseline.

\subsubsection{Worked Example}
\label{sec:worked}

To make the definitions concrete, we hand-evaluate every metric on four
short sequences (a perfectly alternating and a clumped sequence at $n=2$;
a perfectly rotating and a tied sequence at $n=3$), full derivations in
Supplementary Section~S1. Two findings generalise. AALT is the most sensitive
primary ALT metric to same-agent clumping, since it credits a window only for
an exact single win, so back-to-back wins by one agent forfeit that credit
entirely while CALT and EALT still register graded credit. And even a
perfectly rotating finite sequence does not reach $1.0$ on RS when an agent's
wins fail to bookend the run, a disclosed finite-$\nu$ boundary artefact that
vanishes as $\nu$ grows (every experiment below runs $\nu \geq 1{,}000$
episodes).

\subsubsection{Criterion Validity: Synthetic Verification Independent of Learned Behaviour}
\label{sec:criterion-validity}

A natural methodological concern for any new coordination metric is
circularity: if a metric's validity rests only on showing that learned
policies underperform on it, the argument risks being self-referential,
since outcome-based metrics would be judged inadequate because Q-learning
underperforms random play on ALT, yet that underperformance is itself
established only via the ALT metrics. A related
study~\cite{papadopoulos26} addresses this for the ALT family with a
closed-form theoretical null (a birthday-problem-style expression for the
probability of an exclusive winner under uniform random play) whose
predicted CALT floor matches measured random baselines almost exactly. RP's
per-agent, hard-cutoff structure resists a comparable clean derivation, so
we instead validate it with four constructed scenarios of known ground
truth, none involving Q-learning or any learning process; RP's validity
therefore rests on hand-built sequences whose properties we control, not on
the learned behaviour it is later used to evaluate.

Four hand-built ground-truth constructions validate RP independently of any
learning process (full figures and derivations in Supplementary Section~S2):
a positive control (a hand-coded perfect round-robin at the exact episode
counts used below reproduces $\overline{\mathrm{RP}}$ within $1.3\times
10^{-5}$ of $1$ at $n=10$, confirming the boundary artefact of
Section~\ref{sec:worked} vanishes at real scale); a check confirming the ALT
family's population-exclusion calibration~\cite{papadopoulos26} does not
transfer to RP, since RP's per-agent targets are calibrated to the full
population, so excluding agents makes the remainder appear to over-win
rather than to alternate perfectly; a collision-channel sweep separating the
exclusive-win and all-reaches definitions cleanly; and a ballistic
two-agent sweep confirming the weighted RP variants correctly distinguish a
regular-but-unequal allocation from an irregular one. The construction most
directly relevant to the real findings below is reported in full here.

\begin{figure}[t]
\centering
\includegraphics[width=1.0\linewidth]{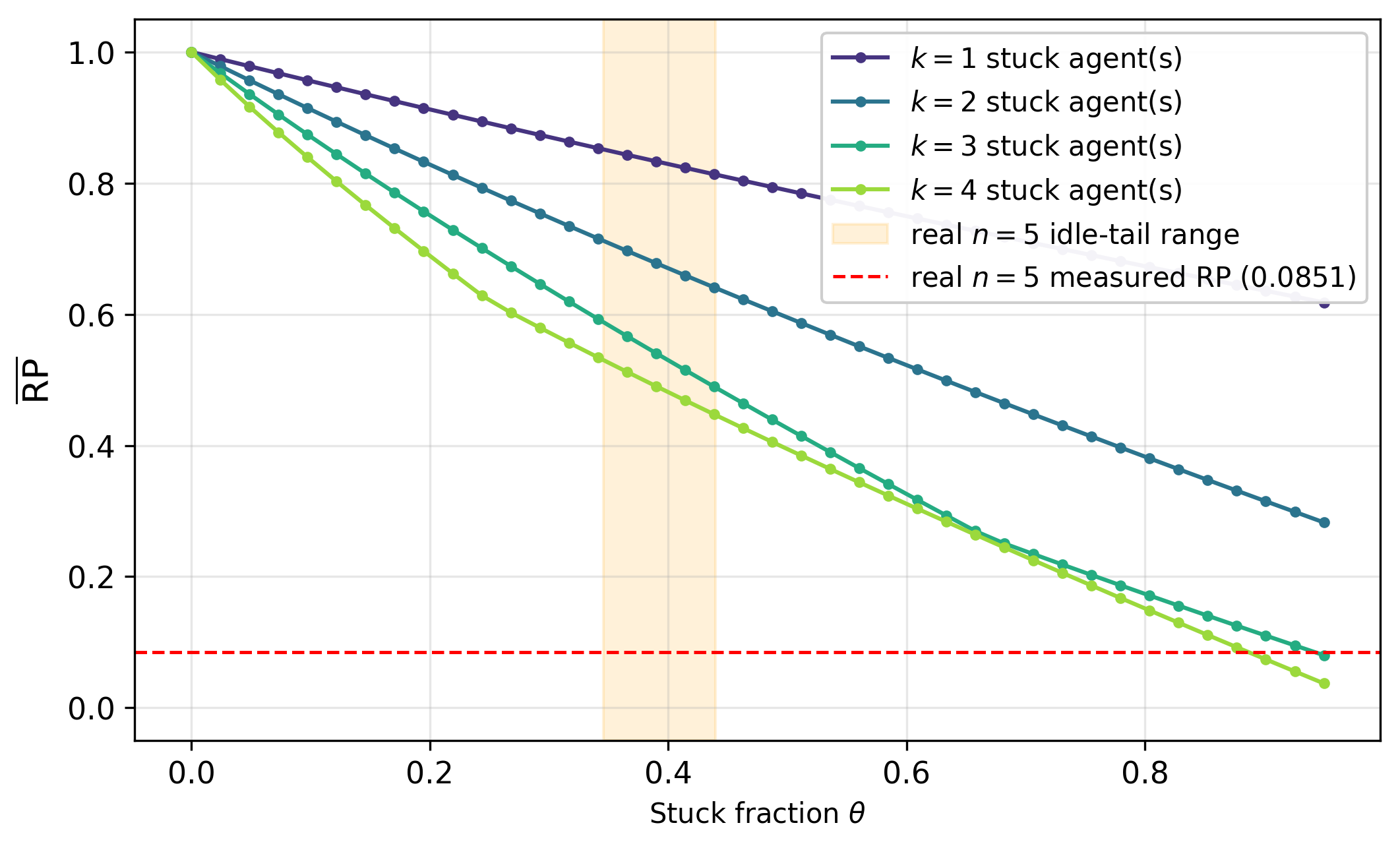}
\caption{Permanent stuck-agent sweep ($n=5$, $\nu=31{,}839$): $\overline{\mathrm{RP}}$
  against the stuck fraction $\theta$, for $k=1,\ldots,4$ permanently idle
  agents. The horizontal line marks the real measured Type-A/ILF
  $\overline{\mathrm{RP}}$; the shaded band marks the real idle-tail range.}
\label{fig:exp2}
\end{figure}

\textbf{Graded degradation: a permanent regime shift.}
The real experiments show Q-learning agents that stop winning exclusively
partway through training and never recover (Section~\ref{sec:experiments}); we
reproduce this directly with $k$ of $n=5$ agents winning normally for a
fraction $1-\theta$ of $\nu=31{,}839$ episodes and then going permanently idle
for the rest, their slots absorbed by round-robin rotation among the remaining
agents. Both sub-measures decrease monotonically in $\theta$ for every $k$,
with RS degrading somewhat faster than WPE throughout (e.g.\ at $k=3$,
$\theta \approx 0.90$: $\mathrm{RS}=0.161$ against $\mathrm{WPE}=0.059$,
$\overline{\mathrm{RP}}=0.110$), and with $k=3$--$4$ stuck agents the combined
$\overline{\mathrm{RP}}$ curve crosses the real measured
$\overline{\mathrm{RP}} = 0.085$ (Type-A, ILF, $n=5$) at $\theta \approx
0.9$--$0.95$ (Figure~\ref{fig:exp2}), and the empirical coordination failure is
quantitatively consistent with most of the population having effectively
abandoned exclusive competition for nearly the entire run, matching the
window-based exclusive-win collapse visible in the training-progression
figures of that study~\cite{papadopoulos26}.

Across all four constructions, RP behaves exactly as its definition predicts
on sequences whose properties are fixed by hand, before any learning algorithm
is involved. The empirical coordination failure reported in
Section~\ref{sec:experiments} is therefore not an artefact of comparing the
metric against itself. RP's response to known ground truth is established
independently, and only then applied to Q-learning.

\textbf{Robustness under an alternative reward rule.} We chose the
population-based ILF/IQF rule as primary a priori, before examining any
results, for measurement cleanliness (full justification in Supplementary
Section~S3). As a complementary check using real Q-learning, we re-ran the
pipeline under the canonical Rosenthal congestion-game payoff~\cite{rosenthal73}
(equal per-claimant shares $r_\mathrm{high}/k$) across the full tested range
$n \in \{2,3,5,8,10\}$. For $n \geq 3$ the resulting RS, WPE,
$\overline{\mathrm{RP}}$, and CALT all fall within $1$--$2\%$ of the ILF
values, narrowing further as $n$ grows (CALT's relative change shrinks
monotonically from $-9.8\%$ at $n=3$ to $-0.2\%$ at $n=8$ and effectively
$0\%$ at $n=10$), so the coordination gap is not an artefact of the
reward-threshold choice. At $n=2$, where every collision is necessarily an
all-agent collision, both CALT and $\overline{\mathrm{RP}}$ drop substantially
under the equal-split rule (CALT: $0.315 \to 0.230$; $\overline{\mathrm{RP}}$:
$0.606 \to 0.325$), a genuine, fully explained, and separately confirming
difference.

\subsubsection{Analytical Comparison of ALT and RP}
\label{sec:analysis}

\emph{Asymptotic Complexity.}

Let $\nu$ denote total episodes and $n$ the agent count. ALT requires
$(\nu - n + 1)$ sliding-window passes, each examining $n$ entries. Its time
complexity is therefore $\Theta(\nu n)$. RP performs a single sweep of the episode
array and constant per-agent post-processing, yielding $\Theta(\nu + n)$.
For the typical regime $\nu \gg n$ the speedup factor approaches $n$, since
$(\nu n) / (\nu + n) \approx n$. In practice we observe speedups of $6$--$20\times$
(see Section~\ref{sec:experiments}), growing with $n$ as the theory predicts.
The constant factor sits above the theoretical minimum because even one ALT
score pays for the whole sliding-window extraction, on top of per-batch Python
function call overhead.

\emph{Sensitivity and Expressiveness.}

The two families differ in what temporal patterns they can detect:

\textit{Alternation within windows (ALT advantage).}
ALT evaluates \emph{which} agents win within each $n$-episode window. It can
therefore distinguish, for example, between a scenario where one agent
monopolises all wins in early windows and a scenario where wins are
spread evenly from the start. RP, working with per-agent statistics
aggregated over the entire sequence, is less sensitive to this temporal structure.

\textit{Rhythmic consistency (RP advantage).}
RP directly measures how close each agent's inter-win gap is to the ideal
$n-1$. This captures scenarios where wins are evenly distributed over time
but come in irregular bursts; such a pattern may score moderately on CALT
but poorly on RS.

\textit{Multi-agent scaling (RP advantage).}
The measured speedup grows with $n$ across the tested range
(Table~\ref{tab:computation_times}), and since ALT costs $O(\nu n)$ against
RP's $O(\nu + n)$, this advantage cannot shrink for populations beyond those
tested here.

\emph{Complementarity and Practical Guidance.}

The two families are complementary. CALT provides rich, window-level
discrimination and is the primary choice for detailed coordination analysis in
small systems. RP provides an efficient, always-computable
signal suitable for large-scale simulations, real-time monitoring, and
preliminary screening, with an advantage that is measured to grow within the
tested range and guaranteed by the complexity separation above not to shrink
beyond it. In systems where both are tractable, using
$\overline{\mathrm{RP}}$ to screen for coordination failure and the three
primary ALT metrics (CALT, EALT, AALT) to diagnose it, monopoly, insufficient
exclusivity, irregular access, or low per-window coverage respectively, gives
stronger diagnostic coverage than either alone. AltRatio $\approx
\sqrt{\mathrm{CALT}}$ maps approximately to the PA-equivalent fraction (the
intercept is near-zero~\cite{papadopoulos26}); for EALT and AALT the mapping
is direct. The auxiliary variants (FALT, qFALT, qEALT) provide additional
discrimination near-perfect alternation for fairness audits distinguishing
dominance from exclusivity. For research contexts comparing multiple agent
strategies, we recommend reporting CALT, EALT, and AALT alongside the
Coordination Score against a random baseline when computation permits, and
$\overline{\mathrm{RP}}$ alone, supplemented by traditional $E$ and RF to
confirm genuine coordination rather than a structural artefact, once a large
agent population or real-time constraint rules ALT out.

\subsubsection{Experimental Setup}
\label{sec:setup}

We run all experiments under the episode scaling formula derived from the state-space
complexity of HJG~\cite{papadopoulos26}:
\[
  \nu = B \cdot \left(\frac{n}{2}\right)^2 \cdot \left(1 + \ln\frac{n!}{2!}\right),
  \quad B = 1000.
\]
This yields the episode counts $\nu = 1{,}000, 4{,}721, 31{,}839, 174{,}583,
385{,}281$ for $n=2,3,5,8,10$ respectively (also listed against their
computation times in Table~\ref{tab:computation_times} below), with
training completing in roughly 5 minutes at $n=2$ up to roughly 20 hours at
$n=10$. For each agent count and reward type (ILF and IQF) we run the
Q-learning experiment for both Type-A and Type-B states. Random baselines
use a fixed budget of $10{,}000$ episodes each. All experiments were run on
an Intel Xeon E5-2640 v4 server (20 cores, 32 GB RAM) running Ubuntu 22.04;
computation-time measurements are single-threaded.

\subsection{Statistical Analysis}
\label{sec:statistical-analysis}

We examine Spearman rank correlations between all RP sub-metric variants
and the three ALT primary metrics across $N=30$ configurations (20
Q-learning runs: $n \in \{2,3,5,8,10\}$, 2 state types, 2 reward types; plus
10 random baselines). Asymptotic standard errors do not exceed $0.070$ for
any entry with $\rho_S \geq 0.929$. We additionally compute partial Spearman
correlations controlling for $n$ to rule out a shared monotonic dependence
on population size, and examine correlation within each fixed agent count
separately. All submetrics use the boundary-inclusive waiting-period
convention (Section~\ref{sec:rp}) and the corrected AALT definition.

\section{Results and Discussion}
\label{sec:results}

\subsection{Q-Learning versus Random Policy: Detecting Coordination Failure}
\label{sec:experiments}

Table~\ref{tab:main_results} reports the core metrics for Type-A states with ILF
rewards, averaged across both random seeds. The pattern is consistent across all
configurations (Type-B, IQF) and we report the full data in the Supplementary
Materials (Table~S4).

\begin{table}[t]
\caption{Metric Values for Q-Learning and Random Policies (Type-A, ILF)}
\label{tab:main_results}
\footnotesize
\centering
\setlength{\tabcolsep}{3pt}
\begin{tabular}{ccccccccc}
\toprule
 & \multicolumn{4}{c}{\textbf{Q-Learning}} & \multicolumn{4}{c}{\textbf{Random}} \\
\cmidrule(r){2-5}\cmidrule(l){6-9}
$n$ & $E$ & RF & $\overline{\mathrm{RP}}$ & CALT & $E$ & RF & $\overline{\mathrm{RP}}$ & CALT \\
\midrule
2  & 0.666 & 0.490 & 0.606 & 0.315 & 0.818 & 0.972 & 0.755 & 0.486 \\
3  & 0.517 & 0.921 & 0.195 & 0.134 & 0.866 & 0.972 & 0.633 & 0.359 \\
5  & 0.457 & 0.963 & 0.085 & 0.059 & 0.727 & 0.954 & 0.456 & 0.243 \\
8  & 0.402 & 0.993 & 0.028 & 0.025 & 0.526 & 0.893 & 0.261 & 0.147 \\
10 & 0.409 & 0.989 & 0.014 & 0.016 & 0.443 & 0.913 & 0.187 & 0.111 \\
\bottomrule
\end{tabular}
\end{table}

Several observations stand out. First, Reward Fairness is \emph{high} ($> 0.9$) for
Q-learning agents with $n \geq 3$, which could na\"ively be interpreted as
successful fair coordination. However, RP and all three primary ALT metrics (CALT,
EALT, AALT) tell a different story. These values are substantially \emph{lower} for
Q-learning than for the random baseline, indicating that the agents have not learned
to coordinate meaningfully.

Second, coordination scores are negative across the board, as shown in
Table~\ref{tab:coordination_scores}. The Q-learning agents systematically
fail to do better than chance at temporal fair division.

\begin{table}[t]
\caption{Coordination Scores (Type-A, ILF): all entries negative, Q-learning
  coordinates \emph{worse than chance}.}
\label{tab:coordination_scores}
\small
\centering
\begin{tabular}{ccccc}
\toprule
$n$ & CS(RP) & CS(CALT) & CS(EALT) & CS(AALT) \\
\midrule
2  & $-61.2\%$ & $-33.2\%$ & $-37.1\%$ & $-24.5\%$ \\
3  & $-119.6\%$ & $-34.9\%$ & $-75.7\%$ & $-31.1\%$ \\
5  & $-68.3\%$ & $-24.3\%$ & $-54.7\%$ & $-20.8\%$ \\
8  & $-31.5\%$ & $-14.3\%$ & $-29.6\%$ & $-10.0\%$ \\
10 & $-21.3\%$ & $-10.7\%$ & $-20.7\%$ &  $-6.6\%$ \\
\bottomrule
\end{tabular}
\end{table}

The magnitude of the coordination gap decreases as $n$ grows, but for a counter-intuitive
reason. The random \emph{baseline} itself improves. With many agents, purely random
resource allocation approximates uniform share distribution by the law of large numbers,
so the random policy naturally approaches the PA ideal at large $n$, raising the bar
that Q-learning must clear. The gap peaks at $n = 3$ ($-120\%$ on RP), where random
agents already alternate reasonably well by chance yet Q-learning still fails to coordinate,
and shrinks to $-21\%$ at $n = 10$ as both policies approach similar low-frequency regimes.
At $n = 2$, Q-learning converges to a monopoly (one agent always wins), so
$\overline{\mathrm{RP}} = 0.606$ is lower than random ($0.755$) and RF drops to $0.49$;
here traditional metrics also detect the failure, making $n = 2$ a distinct case.

The absolute magnitudes of CS(RP) and CS(CALT) differ substantially (e.g.\ at
$n=3$, $-119.6\%$ against $-34.9\%$); this reflects differing random-baseline
floors, not an inconsistency between the two families. RP's validated claim is
ordinal, not cardinal. Table~\ref{tab:correlations} establishes near-perfect
rank correlation ($\rho_S \geq 0.95$) between RP and every ALT variant, meaning
the two families agree on which configurations coordinate better or worse,
which is what a scalable substitute requires, not agreement on scale. Here the
random baseline itself reaches $\overline{\mathrm{RP}} = 0.633$ at $n=3$ but
only $\mathrm{CALT} = 0.359$. RS and WPE's per-agent tolerance bands score an
unstructured random arrival process comparatively leniently, while CALT's
windowed, population-joint construction is more exacting about simultaneous
arrivals. A higher random floor mechanically inflates $|\mathrm{CS}|$ for a
similar absolute drop, since the numerator $M_\mathrm{QL} - M_\mathrm{rand}$
grows more negative while the denominator $1-M_\mathrm{rand}$ shrinks. This is
not a defect to be tuned away. RS and WPE's weights are fixed by Perfect
Alternation's own structural requirements (Section~\ref{sec:rp}), not chosen
to match CALT's numerical scale, and forcing such a match would overfit to
this dataset's particular tie structure while destroying the closed-form,
PA-grounded interpretation that makes RP usable at population sizes where CALT
cannot be computed at all.

\subsection{RS Replaces AWE: Eliminating Collapse and Asymmetry}
\label{sec:rs-replaces-awe}

In an earlier conference paper~\cite{papadopoulos25rp}, the rhythm sub-measure was
AWE with a hard threshold at $\bar{r}_i = 2(n-1)$. In the present experiments,
$\overline{\mathrm{AWE}} = 0$ for \emph{all} configurations with $n \geq 3$
(both Q-learning and random), because agents win infrequently and their mean
gaps $\bar{r}_i$ exceed $2(n-1)$ by large margins. Under AWE, this reduces
$\overline{\mathrm{RP}}$ to $\overline{\mathrm{WPE}}/2$, losing the rhythmic
dimension entirely.

RS eliminates this collapse. Even when $\bar{r}_i \gg r_i^*$, the RS value is
$r_i^*/\bar{r}_i$, which remains positive and decreasing, with a larger gap
yielding a lower (not zero) score and an everywhere non-zero derivative. The only case
where RS $= 0$ is $k_i < 2$ (agent wins at most once), which correctly
represents total coordination failure. This continuous behaviour also means
that for configurations where $n = 10$ and $\bar{r}_i \approx 30 \cdot (n-1)$,
the RS value is $\approx 1/30$ rather than $0$, preserving the rank ordering
across configurations that AWE would flatten.

Crucially, RS is also \emph{symmetric}. The former AWE formula
$1 - |\bar{r}_i - r_i^*|/r_i^*$ is linear in the deviation and clips over-winning
(small $\bar{r}_i$) at the same rate as under-winning (large $\bar{r}_i$)
only within the feasible range; beyond $2r_i^*$, under-winning is silently
set to $0$ regardless of degree. RS assigns equal scores to $\bar{r}_i = c \cdot r_i^*$
and $\bar{r}_i = r_i^*/c$ for any $c > 1$, satisfying the natural requirement
that deviating twice as fast in either direction is equally bad. In short,
the AWE collapse demonstrated above is fully resolved by the RS sub-measure
introduced here.

\subsection{Scalability: Computation Time Analysis}
\label{sec:scalability}

Table~\ref{tab:computation_times} reports wall-clock computation times for RP
versus the full ALT family across all agent counts with their corresponding episode
budgets.

\begin{table}[t]
\caption{Computation Time: One RP Score vs.\ One ALT Score, Single Thread
  (Type-B states, population-based reward). ALT is the mean over the six
  variants, with their range; per-variant timings are in Supplementary
  Table~S5.}
\label{tab:computation_times}
\footnotesize
\centering
\setlength{\tabcolsep}{3pt}
\begin{tabular}{cccccc}
\toprule
$n$ & Episodes $\nu$ & RP (s) & ALT mean (s) & ALT range (s) & Speedup \\
\midrule
2  & 1{,}000   & 0.0019 & 0.0114 & 0.0098--0.0137 & $6\times$ \\
3  & 4{,}721   & 0.0039 & 0.0371 & 0.0327--0.0443 & $9\times$ \\
5  & 31{,}839  & 0.0307 & 0.437  & 0.400--0.511   & $14\times$ \\
8  & 174{,}583 & 0.2450 & 4.54   & 4.29--5.15     & $19\times$ \\
10 & 385{,}281 & 0.6733 & 13.19  & 12.55--14.86   & $20\times$ \\
\bottomrule
\end{tabular}
\end{table}

\begin{figure}[t]
\centering
\includegraphics[width=1.0\linewidth]{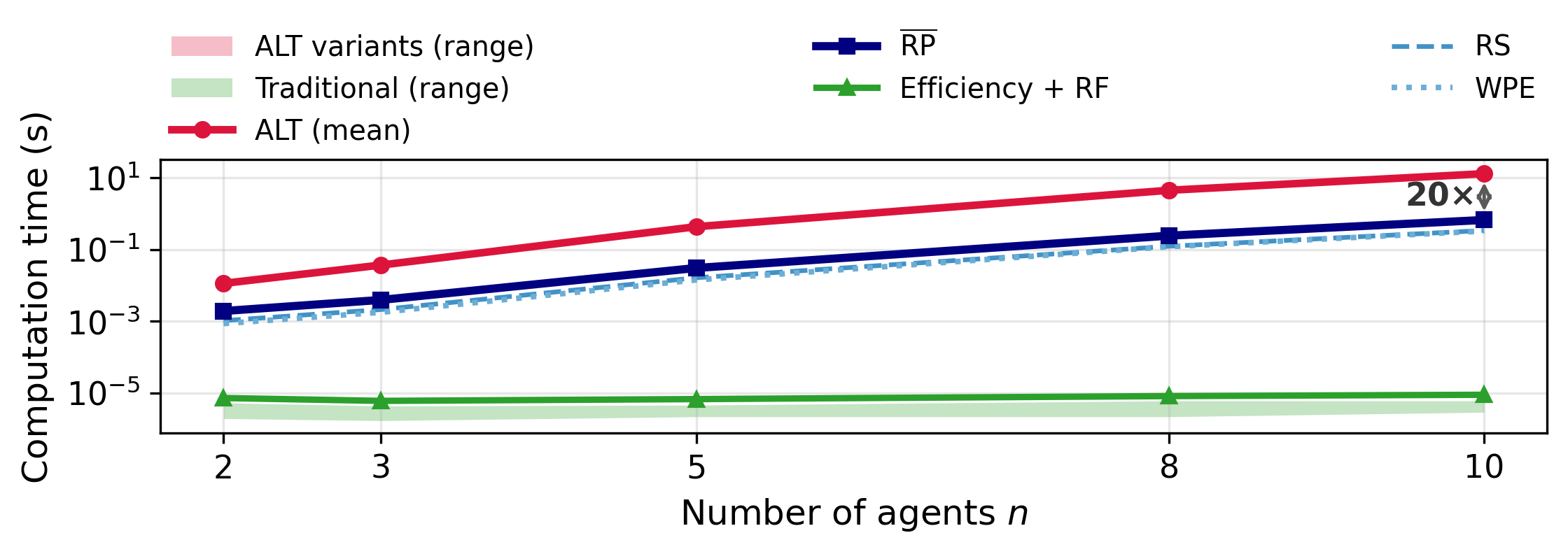}
\caption{Single-thread computation time, log scale, from the run of
  Table~\ref{tab:computation_times}. Each line is a family's mean and each
  band its range; the ALT band is narrow because the six variants cost
  almost the same, for the reason given with that table. The rhythm line
  carries the stored AWE timing, which is what RS costs too, the two
  differing only in a final closed-form step.}
\label{fig:computation_times}
\end{figure}

The speedup grows with $n$, from $6\times$ at $n=2$ to $20\times$ at $n=10$,
consistent with the $O(n)$ prediction for the ratio $\nu n / (\nu+n) \approx n$.
It exceeds that ratio because ALT's work per window entry is about twice RP's
per episode. That work is the sliding-window extraction, which the six variants
share, and which is why they cost so nearly the same as one
another (Supplementary Table~S5). Growth across this range is slightly
sublinear, and $\nu$ is fixed by $n$ here, so these five points cannot say whether
the agent count or the episode budget is responsible. Timing the metrics
directly on synthetic sequences separates them and needs no training, so it
reaches populations no learning run here can. Varying $\nu$ alone over a
$400$-fold range moves the speedup only within $7.98$ to $12.31$, without
trend, while holding $\nu = 100n$ takes it from $8.0\times$ at $n=10$ to
$1220\times$ at $n=2000$, proportional to $n$ throughout (Supplementary
Section~S7). The constant of proportionality
stays between $0.577$ and $0.657$ for $n \geq 50$, and Supplementary
Section~S7 reports what happens once $\nu$ stops being well above $n$. Figure~\ref{fig:computation_times} shows the tested range on a
logarithmic axis, each family's spread shaded behind its mean.

At $n = 10$ one RP score takes under a second against $13$ for one ALT score,
and for larger $n$ the episode count grows faster than linearly (the $n!$
factor of Section~\ref{sec:setup}), so ALT's $O(\nu n)$ cost compounds on two
fronts while RP's $O(\nu+n)$ tracks $\nu$ alone. The gap can only widen.

\subsection{Metric Correlation Analysis}
\label{sec:correlation-analysis}

\begin{table}[t]
\caption{Spearman $\rho_S$, combined $\overline{\mathrm{RP}}$ vs.\ the ALT
  primary metrics ($N=30$; all entries $p<0.001$). The 8 individual
  sub-metric rows this table summarises are in Supplementary Table~S3.}
\label{tab:correlations}
\small
\centering
\begin{tabular}{cccc}
\toprule
Combined $\overline{\mathrm{RP}}$ variant & CALT & EALT & AALT \\
\midrule
RS$_\text{excl}$ $+$ WPE$_\text{excl}$   & $0.970$ & $0.996$ & $0.967$ \\
RS$_\text{reach}$ $+$ WPE$_\text{excl}$  & $0.975$ & $0.976$ & $0.963$ \\
RS$_\text{excl}$ $+$ WPE$_\text{reach}$  & $0.960$ & $0.980$ & $0.956$ \\
RS$_\text{reach}$ $+$ WPE$_\text{reach}$ & $0.990$ & $0.954$ & $0.971$ \\
\bottomrule
\end{tabular}
\end{table}

We do not seek a single "best" submetric here. RS, WPE, AWE and the raw
avg\_wait quantity are all facets of the same lightweight RP family, and the
goal of Table~\ref{tab:correlations} is to show that this \emph{entire family},
under either win-event definition, tracks the expensive ALT family closely
enough to serve as its economical substitute (full per-row discussion,
including the avg\_wait sign flip between exclusive-win and all-reaches
definitions and the AWE$_\text{excl}$/AWE$_\text{reach}$ asymmetry, in
Supplementary Section~S4). All four excl/reach combinations of RS and WPE
achieve $\rho_S \geq 0.95$ against every ALT metric, clustering within a
narrow $0.95$--$0.99$ band regardless of which win-event definition is used
for which component, so a user need not resolve that choice precisely to
obtain a reliable RP score. This alignment is not an artifact of joint
scaling with $n$: partial Spearman correlations controlling for $n$ remain
strong ($\rho_S = 0.941$ CALT, $0.993$ EALT, both $p < 10^{-8}$, $N=30$), and
within each fixed agent count $n \geq 3$, $\overline{\mathrm{RP}}$ and CALT
correlate at $\rho_S \approx 0.94$ uniformly. The exception is $n=2$, where
Q-learning converges to monopoly and the few distinct configurations make
rank correlation degenerate; $n=2$ is treated as a separate regime throughout.

Although the sample spans only five distinct agent counts, the monotone relationship
is consistent across all sub-groups (Q-learning vs.\ random, Type-A vs.\ Type-B,
ILF vs.\ IQF). The correlations between $\overline{\mathrm{RP}}$ and traditional
metrics are substantially weaker ($r \approx 0.31$ with Efficiency, $r \approx 0.18$
with Reward Fairness), confirming that RP captures a distinct temporal dimension
not reflected in standard metrics.

This pattern holds across all four Q-learning configurations against their
respective random baselines, not only the Type-A/ILF condition tabulated
above (Supplementary Figure~S4).

\emph{RS as a scalable proxy for CALT.}
RS$_\text{excl}$ can be viewed as a lightweight approximation of CALT's per-agent
coordination signal. Rather than computing $O(\nu n)$ sliding-window batch scores,
RS estimates each agent's proximity to the PA gap target in $O(k_i)$ time and
combines results into $\overline{\mathrm{RP}}$ in $O(\nu + n)$ overall. Despite
this simplicity, RS$_\text{excl}$ tracks CALT at $\rho_S = 0.959$ and EALT at
$\rho_S = 0.986$; the combined $\overline{\mathrm{RP}}$ (RS$_\text{excl}$ +
WPE$_\text{excl}$) achieves $\rho_S = 0.970$ (CALT), $0.996$ (EALT), and $0.967$
(AALT). In other words, the fine-grained window scoring of CALT does not reveal
substantially different coordination failures from what the $O(\nu + n)$ RP family
already detects, at least within the $n \leq 10$ regime studied here.

\subsection{Summary of Empirical Findings}
\label{sec:summary}

\begin{enumerate}
  \item \textbf{Coordination failure is widespread}, $3$--$120\%$ below
    random on RP and $7$--$35\%$ on CALT across all tested configurations
    (peaking at $n=3$).
  \item \textbf{Traditional metrics are misleading}: Reward Fairness exceeds
    $0.9$ even when coordination scores are strongly negative.
  \item \textbf{The RP family economically substitutes for ALT}: every
    excl/reach pairing achieves $\rho_S \geq 0.95$ against every ALT metric
    ($N=30$), validating the entire lightweight family, not one designated
    formula.
  \item \textbf{RP scales gracefully}, reaching a $20\times$ speedup at
    $n=10$, a margin the complexity separation guarantees cannot shrink.
  \item \textbf{RS provides a continuous, symmetric rhythm signal}, unlike
    AWE, which collapses to zero for $n \geq 3$.
\end{enumerate}

\subsection{Implications for Temporal Fair Division Theory and Mechanism Design}
\label{sec:discussion}

Our results show learning agents producing \emph{worse} outcomes than random
resource competition, falling further from Perfect Alternation (the temporally
fair solution) than chance itself, a counterintuitive
finding explained by the asymmetry of exploration: $\varepsilon$-greedy
Q-learning cycles through policies with occasional wins during early
training without discovering the correlation structure turn-taking
requires, then locks into locally optimal but globally suboptimal
strategies as $\varepsilon$ decays, a manifestation of the well-known
coordination failure in multi-agent Q-learning~\cite{shoham08}. From a fair
division perspective, the PA regime is not self-enforcing under independent
learning, so mechanism design interventions may be required. Supplementary
Section~S6 sketches how RP and ALT might serve as optimisation objectives
rather than only as diagnostics, and why a metric used as a training signal
stops being an independent measure of the behaviour it shaped.

\subsection{Limitations}
\label{sec:limitations}

The current framework has three principal limitations. First, we base primary
conclusions on Type-A for state-space compactness; the Type-B winner-flag
encoding was verified programmatically (all primary conclusions nonetheless
rest on Type-A). Second, the episode budget formula assumes identical complexity scaling
across all configurations; for very large $n$ this may underestimate the
effective state-space size.

Third, every learning result here uses independent tabular Q-learning, for the
interpretability reasons given in Section~\ref{sec:setup}. Whether centralised
training, function approximation, or explicit fairness objectives narrow the
coordination gap is open, and is the natural next study. This bounds the
empirical claim rather than the metrics: RS, WPE and the ALT family read only
the sequence of episode winners, so they transfer unchanged to any learner,
and to scripted or human play.

\section{Conclusion}
\label{sec:conclusion}

We have presented a framework for measuring temporal fairness in repeated
multi-agent resource competition, grounded in fair division theory through the
connection between Perfect Alternation and temporal proportionality and
envy-freeness. It spans three layers: traditional metrics (Efficiency, Reward
Fairness) capturing aggregate outcomes; the ALT family capturing detailed
coordination quality; and the Rotational Periodicity family, linear-time
metrics decomposing temporal fairness into rhythmic and distributional
dimensions. Empirically, Q-learning agents consistently fail to achieve
temporal fair division, performing worse than random baselines on RP and all
three primary ALT metrics while maintaining high Reward Fairness, a failure
invisible to traditional metrics alone, with RP achieving $6$--$20\times$
speedups at near-perfect rank correlation ($\rho_S \geq 0.95$).
We hope these results motivate including temporal fairness metrics in
evaluation pipelines for multi-agent resource allocation, since standard
metrics actively conceal coordination failure.

Future directions include: (i) reward-shaping mechanisms based on RP to guide
agents towards temporal fair division; (ii) validating the RS--CALT correlation
for larger populations ($n>10$) where ALT's $O(\nu n)$ cost is expected
impractical; (iii) extending to heterogeneous agents via the weighted RS and
ERP variants; and (iv) real-world scheduling and resource allocation benchmarks.

\section*{Acknowledgment}
The authors disclose the use of Anthropic's Claude as an AI assistant: for
editorial refinement of the authors' own prose, for \LaTeX{} and notation
formatting, and to assist in writing parts of the analysis and visualisation
code, specifically the scripts behind the synthetic validation constructions
of Section~\ref{sec:criterion-validity} and the computation-time measurements
of Figure~\ref{fig:computation_times}. The game environment, the learning
agents, and the metric implementations are the authors' own. No claim, result,
or conclusion here was produced by an AI system; the research questions, the
design of the game and of the metrics, the experimental design, and the
interpretation of all results are the authors' exclusive intellectual
contribution, and they bear full responsibility for the manuscript.

\section*{Author Contributions}
N. Al. Papadopoulos conceived the study, formalised the fair division framework
and all metrics, designed and conducted the experiments and synthetic validation
studies, developed all software (environment, agents, metrics, and analysis
scripts), and drafted the manuscript. M. Sanchez-Fibla contributed to the
original creation of the game. I. T. Freire contributed the idea of
AltRatio-based benchmarking. I. T. Freire, M. Sanchez-Fibla, and K. E. Psannis
supervised the work, critically revised sections of the manuscript, and reviewed
the final draft. All authors read and approved the final manuscript.

\section*{Funding}
The authors did not receive funding for this work.

\section*{Conflicts of Interest}
The authors declare no competing interests.

\section*{Data Availability}
Code implementing the Honey-Jar Game environment, the Q-learning agents, and
all metrics will be made available upon publication. The supplementary
material included below contains the worked-example derivations
(Section~S1), the three ground-truth constructions not plotted here
(Section~S2), the reward-rule design justification (Section~S3), the
per-sub-metric correlations and named RP variants (Section~S4), metric
values for every configuration (Section~S5, Table~S4), and the per-variant
computation times and a scaling study (Section~S7).

\clearpage
\section*{Supplementary Material}
\renewcommand{\thesection}{S\arabic{section}}
\renewcommand{\thetable}{S\arabic{table}}
\renewcommand{\thefigure}{S\arabic{figure}}
\setcounter{section}{0}
\setcounter{table}{0}
\setcounter{figure}{0}

\noindent
This document collects the derivations, constructions, and full result tables
that the main text summarises for space. Nothing here is new material. Every
claim below is stated in condensed form in the main text, which points to the
section, table, or figure of this document where it is worked out in full.
Section~\ref{sup:worked} gives the complete hand-derivations behind the worked
example; Section~\ref{sup:synthetic} gives the three ground-truth constructions
not plotted in the main text; Section~\ref{sup:reward} gives the full design
justification for the primary reward rule; Section~\ref{sup:corr} gives the
per-sub-metric correlation table and its discussion;
Section~\ref{sup:fulltables} gives metric values for every state and reward
configuration; Section~\ref{sup:shaping} sketches the reward-shaping and
mechanism-design question the main text's findings leave open; and
Section~\ref{sup:timings} breaks the computation-time table down by variant and
measures how the cost separation scales to far larger populations.

\section{Worked Example: Full Derivations}
\label{sup:worked}

We evaluate every metric on four short sequences. We write the winner of each
episode as a letter; a brace denotes a simultaneous arrival (a tie).
Table~\ref{tab:worked} collects the results, and the two paragraphs below
derive them by hand.

\begin{table*}[t]
\centering
\caption{All metrics on four illustrative sequences. Efficiency ($E$) and
  Reward Fairness (RF) use ILF rewards. For $n=2$ the perfectly alternating
  and clumped sequences agree on efficiency, reward fairness, and WPE; CALT,
  EALT, AALT, and RS all register the clumping, with AALT dropping furthest
  since it requires an exact single win per sliding window.}
\label{tab:worked}
\small
\begin{tabular}{lcccccccc}
\toprule
Sequence & $E$ & RF & CALT & EALT & AALT & RS & WPE & $\overline{\mathrm{RP}}$ \\
\midrule
$ABABABAB$ \; ($n{=}2$, PA)        & 1.00 & 1.00 & 1.00 & 1.00 & 1.00 & 1.00 & 1.00 & 1.00 \\
$ABBAABBA$ \; ($n{=}2$, clumped)   & 1.00 & 1.00 & 0.68 & 0.79 & 0.57 & 0.78 & 1.00 & 0.89 \\
\midrule
$ABCABC$ \; ($n{=}3$, PA)          & 1.00 & 1.00 & 1.00 & 1.00 & 1.00 & 0.89 & 1.00 & 0.94 \\
$AA\{BC\}AA\{BC\}$ \; ($n{=}3$)    & 0.89 & 0.44 & 0.47 & 0.67 & 0.00 & 0.31 & 0.00 & 0.15 \\
\bottomrule
\end{tabular}
\end{table*}

\paragraph{Two agents ($n=2$, $\nu=8$).}
The perfectly alternating sequence $ABABABAB$ and the clumped sequence
$ABBAABBA$ both award each agent four wins, so both are temporally proportional
and envy-free at every cycle boundary (after episodes $2,4,6,8$ the agents are
tied in cumulative wins). Efficiency, reward fairness, and WPE confirm this
directly. All three equal $1.0$ for both sequences, since none of them looks
at rhythm, only at totals. The rhythm sub-measure RS exposes the difference
between the two sequences. In $ABBAABBA$, agent $A$ wins at episodes $0,3,4,7$;
its waiting periods are the internal gap between $0$ and $3$ ($2$ episodes),
between $3$ and $4$ ($0$ episodes), and the trailing period after episode $7$
($0$ episodes, since $7$ is the final episode), giving $\bar r_A = 2/3$ and
$\mathrm{RS}_A = \bar r_A/r^* = 0.75$. Agent $B$ wins at $1,2,5,6$, with a leading
period before episode $1$ ($1$ episode), internal gaps of $0$ and $2$, and a
trailing period after episode $6$ ($1$ episode), giving $\bar r_B = 4/5$ and
$\mathrm{RS}_B = 0.80$. The resulting $\overline{\mathrm{RP}} = 0.89 < 1$ flags
the irregular rhythm. The ALT family responds too. CALT drops to $0.68$ and
EALT to $0.79$, both from the larger tie-adjacent penalty and lower per-window
agent coverage that windows containing two consecutive wins by the same agent
incur. AALT drops furthest, to $0.57$, since it credits a window only when
exactly one agent registers a solo win in it, so the two windows where an agent
wins twice in a row forfeit that window's AALT credit entirely, whereas EALT's
broader per-window coverage measure still assigns partial credit for reduced
but nonzero reach diversity. RS and the ALT family therefore move together on
this sequence, a small, hand-verifiable instance of the rank correlation
($\rho_S \geq 0.95$ against every ALT variant) established empirically across
the full dataset in the main text. This is exactly the regularity dimension of
Perfect Alternation that proportionality and envy-freeness leave unconstrained,
and which the RS component of RP was built to isolate.

\paragraph{Three agents with ties ($n=3$, $\nu=6$).}
Even the perfectly rotating $ABCABC$ does not score a clean $1.0$ on RS.
Agent $B$ wins at episodes $1$ and $4$, with a leading period of $1$ (episode $0$),
an internal gap of $2$, and a trailing period of $1$ (episode $5$), giving
$\bar r_B = 4/3 < r^* = 2$ and $\mathrm{RS}_B = 2/3$; agents $A$ and $C$, whose
wins bookend the sequence exactly ($0,3$ and $2,5$), score $\mathrm{RS}_A =
\mathrm{RS}_C = 1$. The resulting system average $\mathrm{RS} = 0.89$ is a
finite-sequence boundary artefact, not a coordination flaw. With only $\nu=6$
episodes and three agents, one agent cannot avoid an asymmetric position
relative to the sequence's start and end. This effect vanishes as $\nu$ grows;
every experiment in the main text runs $\nu \geq 1{,}000$ episodes, where such
boundary contributions are negligible.

Now compare $ABCABC$ with $AA\{B,C\}AA\{B,C\}$, where $A$ takes every solo win
and $B,C$ only ever arrive together. Every sliding window of this periodic
sequence has the same composition, two solo wins by $A$ and one tie between
$B$ and $C$, so every batch score equals the run average. The tie penalty
$\sum_k(n-y_k) = (3-1)+(3-1)+(3-2) = 5$ is weighted by the reach-diversity
score $\beta_j^{\mathrm{qFALT}} = (f_j/t_j)^2 = (3/4)^2 = 0.5625$, since all
three agents reach at least once but four terminal arrivals accrue against
them, giving $\mathrm{CALT} = 5 \times 0.5625/6 = 0.47$. EALT, weighted the
same way by reach diversity, gives $(w_j \cdot f_j)/n^2 = (2 \times 3)/9 =
0.67$. Under the exclusive-win definition used by RS and WPE, $B$ and $C$
record no solo victories at all, so each agent's entire $6$-episode run counts
as a single failed waiting period: $\bar r_B = \bar r_C = 6$, giving
$\mathrm{RS}_B = \mathrm{RS}_C = r^*/6 = 1/3$, and $\mathrm{WPE}_B =
\mathrm{WPE}_C = 0$ (zero win events against an ideal of $t^*=2$). Agent $A$'s
clustered wins at episodes $0,1,3,4$ give periods $(0,1,0,1)$, $\bar r_A = 1/2$
and $\mathrm{RS}_A = 1/4$; with $k_A = 4 \geq 2t^*$, $\mathrm{WPE}_A = 0$ as
well. AALT reaches exactly $0.00$ for the same underlying reason as WPE.
Every sliding window contains either two consecutive solo wins by $A$, which
forfeits that window's AALT credit under the exactly-one-win rule, or the
tie between $B$ and $C$, who never win alone at all, so no window ever
credits any agent. The combined $\overline{\mathrm{RP}} = 0.15$ signals
near-total coordination failure, which reward fairness ($0.44$) reflects only
weakly; CALT ($0.47$) and EALT ($0.67$) register the same failure less starkly
still. AALT and WPE reach the absolute floor of $0.00$, and
$\overline{\mathrm{RP}}$ lands between WPE's floor and RS's $0.31$, exactly
matching its definition as their average, so it is not the single most extreme
signal here but the one that combines both dimensions of the failure into one
number.

\section{Synthetic Validation Constructions}
\label{sup:synthetic}

The main text reports the stuck-agent construction in full, since it is the one
that speaks most directly to the empirical findings. The remaining three
constructions are given here. None of them involves Q-learning or any learning
process; each fixes the properties of the sequence by hand and checks that RP
responds as its definition predicts.

\subsection*{Positive control}
\begin{figure}[ht]
\centering
\includegraphics[width=0.78\linewidth]{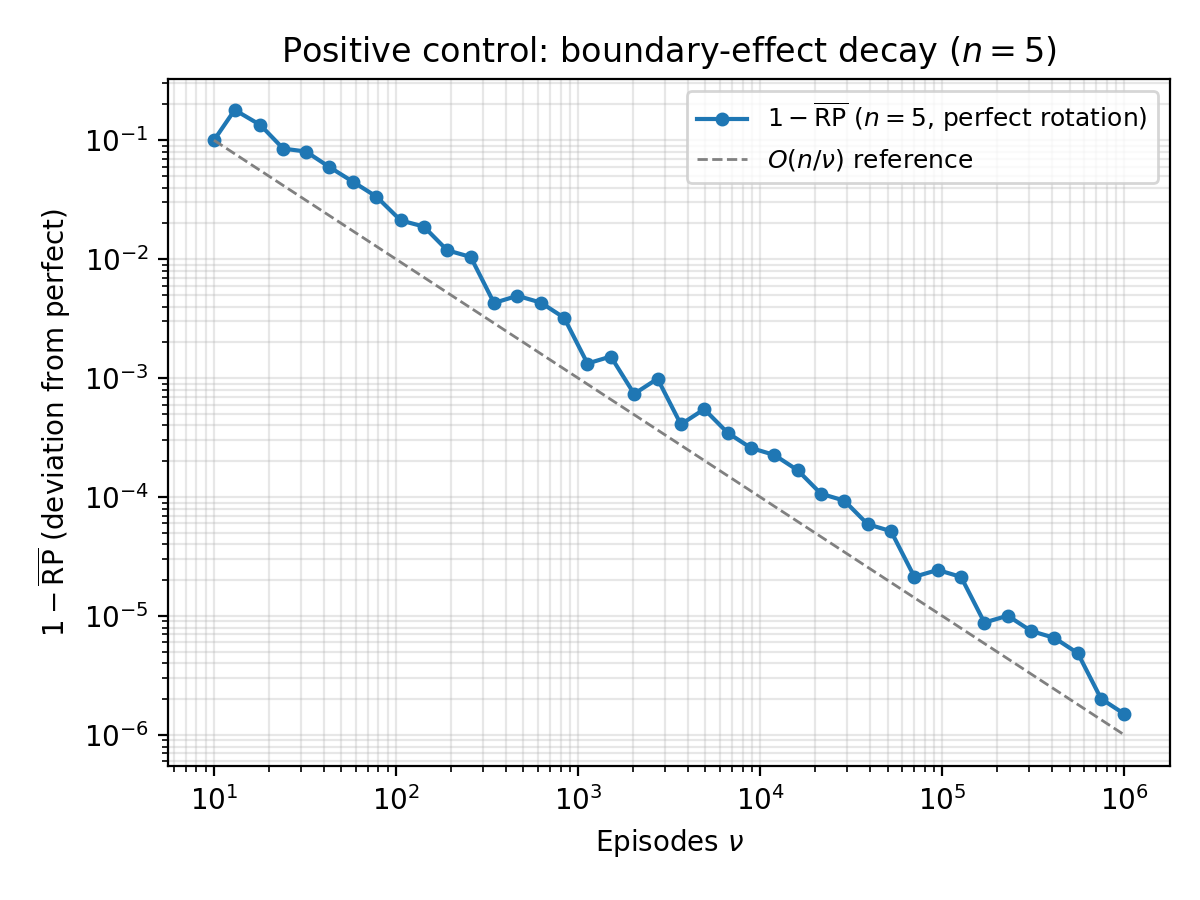}
\caption{Positive control: boundary-effect decay for a hand-coded perfect
  rotation ($n=5$). The deviation $1-\overline{\mathrm{RP}}$ tracks the
  $O(n/\nu)$ reference line, confirming that finite-sequence artefacts vanish
  at the scale used throughout the experimental study.}
\label{fig:exp1}
\end{figure}

Constructing a perfect round-robin at the exact episode counts used in the
experimental study ($\nu = 1{,}000,\ 4{,}721,\ 31{,}839,\ 174{,}583,\ 385{,}281$
for $n = 2,3,5,8,10$) gives $\overline{\mathrm{RP}}$ within $1.3\times 10^{-5}$
of $1$ at $n=10$, with the residual deviation shrinking as $\nu$ grows
(Figure~\ref{fig:exp1}, a log-log sweep of $\nu$ at $n=5$, decaying at the
$O(n/\nu)$ rate that the boundary artefact of Section~\ref{sup:worked}
anticipates).

\subsection*{Why the population-exclusion calibration does not transfer to RP}
The ALT family's graded-response check sweeps synthetic populations in which
$x$ of $n$ agents alternate perfectly while the remaining $n-x$ are entirely
excluded~\cite{papadopoulos26}; this is well-defined for batch metrics because
an empty seat simply lowers window coverage. An earlier conference paper already anticipated that the same construction would not
transfer to RP,
since RP primarily evaluates individual rotation patterns rather than
coordinated alternation~\cite{papadopoulos25rp}. We confirm this concretely.
With $n=4$ and $x=2$ active agents rotating perfectly between themselves while
two are excluded, the two active agents receive $\mathrm{RS}=0.333$ (not $1$)
and $\mathrm{WPE}=0$, because their per-agent targets ($r^*=n-1$, $t^*=\nu/n$)
are calibrated to the full population; excluding agents makes the remaining
ones win faster than that population-wide target, which RS and WPE penalise
symmetrically, exactly as they would penalise monopolisation. We therefore do
not attempt a population-exclusion calibration for RP and instead validate it
with constructions in which every agent remains part of the population
throughout.

\subsection*{Graded degradation: a collision channel}
\begin{figure}[ht]
\centering
\includegraphics[width=0.78\linewidth]{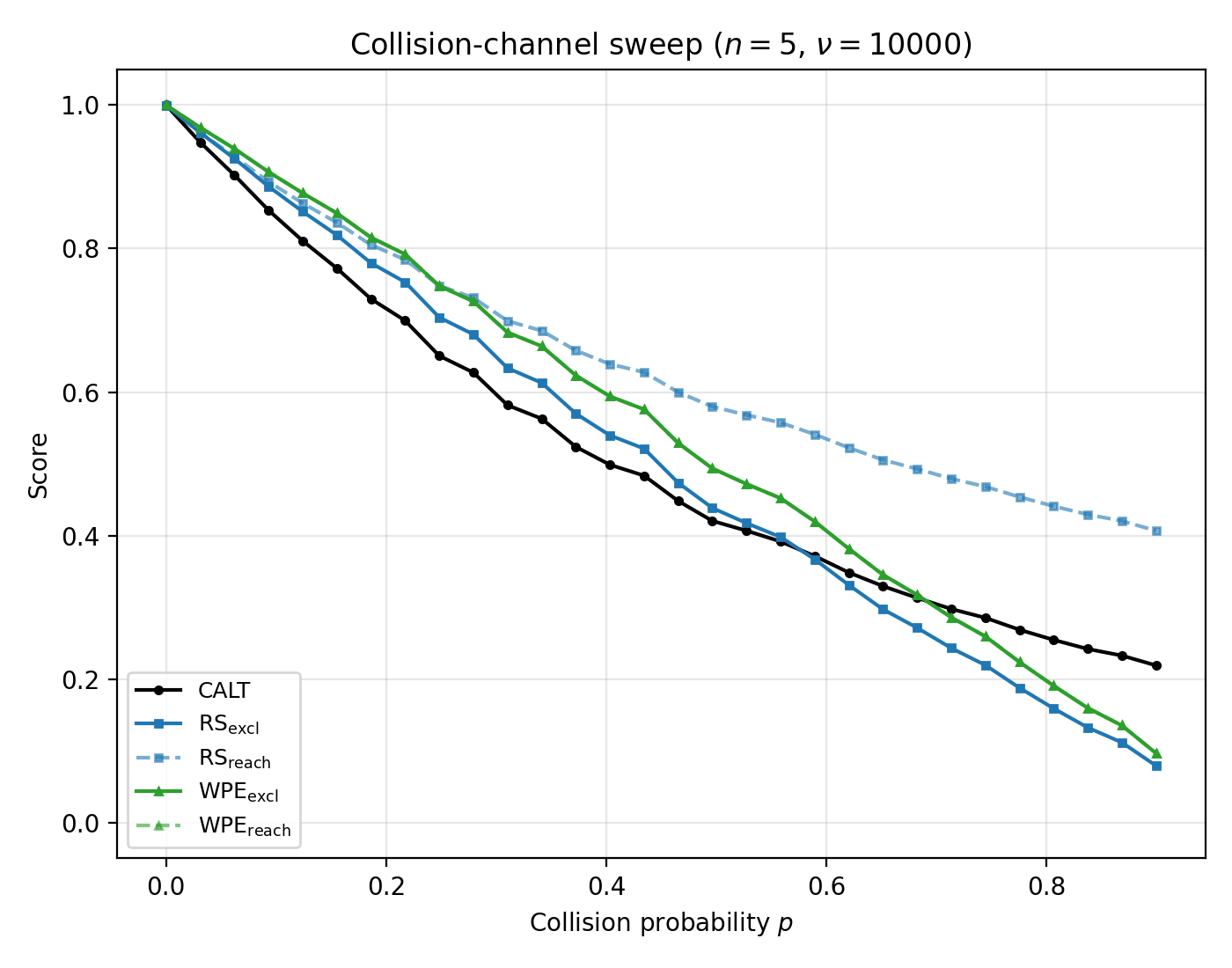}
\caption{Collision-channel sweep ($n=5$, $\nu=10{,}000$): CALT and the
  exclusive/all-reaches RS and WPE variants against the injected collision
  probability $p$.}
\label{fig:exp3}
\end{figure}

A fixed rotation order is corrupted by injecting, at each turn, a collision
with probability $p$ (a randomly chosen second agent joins the intended
winner, turning a clean win into a tie); $p=0$ exactly reproduces the positive
control above. Sweeping $p$ (Figure~\ref{fig:exp3}) separates the exclusive
and all-reaches definitions concretely. CALT and the exclusive-win submetrics
degrade fastest, since a collision destroys an exclusive win outright, while
$\mathrm{RS}_\text{reach}$ degrades most gently, since colliding agents still
count as having reached the terminal. $\mathrm{WPE}_\text{excl}$ and
$\mathrm{WPE}_\text{reach}$ coincide exactly at every $p$ in this construction,
a direct consequence of the symmetric way collisions simultaneously remove an
exclusive win from one agent and add a reach event to another.

\subsection*{Connection to classical equilibrium theory}
\begin{figure}[ht]
\centering
\includegraphics[width=0.92\linewidth]{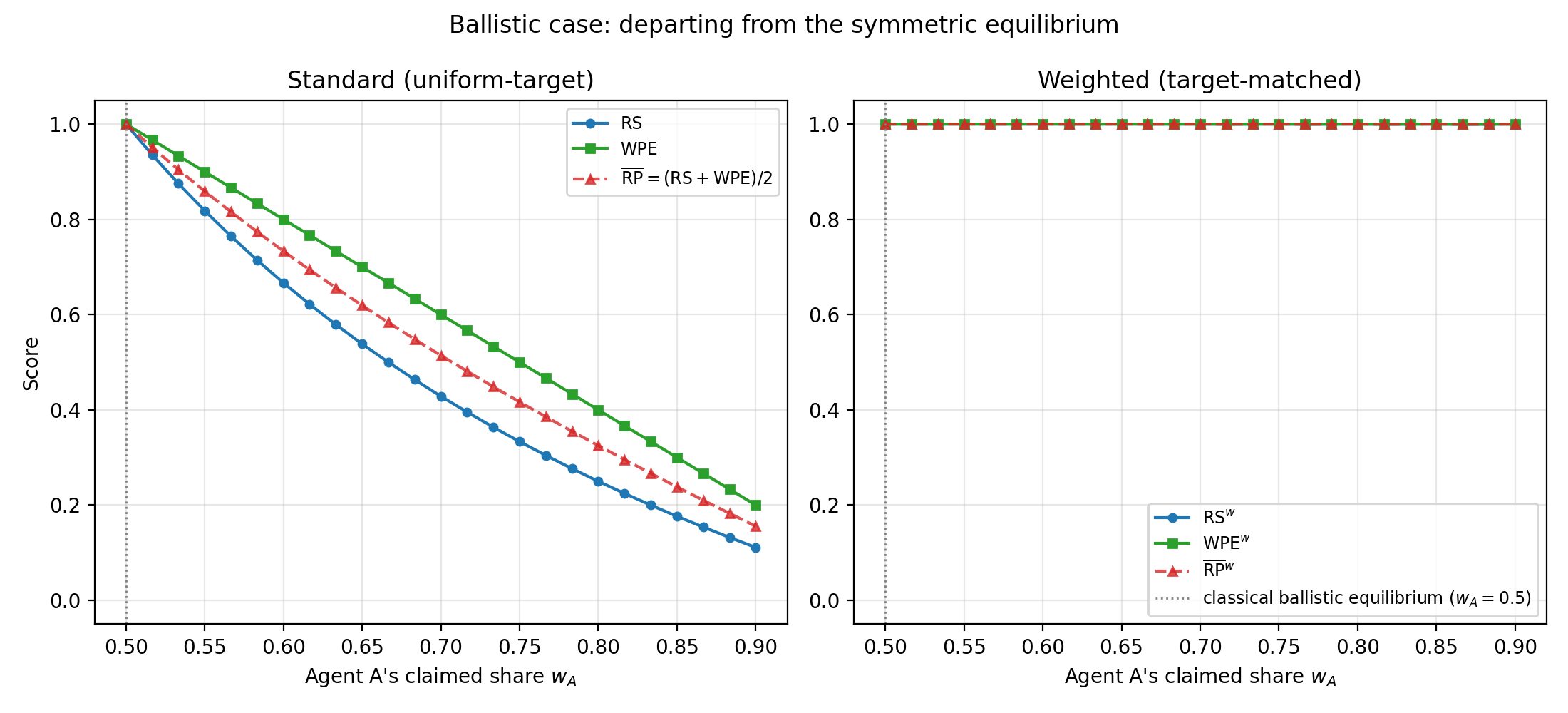}
\caption{Ballistic two-agent case: RS, WPE, and $\overline{\mathrm{RP}}$ (left,
  standard/uniform-target) versus $\mathrm{RS}^w$, $\mathrm{WPE}^w$, and
  $\overline{\mathrm{RP}}^w$ (right, weighted/target-matched) as agent $A$'s
  claimed share $w_A$ departs from the symmetric ($0.5$) equilibrium on an
  otherwise perfectly regular schedule.}
\label{fig:exp4}
\end{figure}

Finally, we tie RP to the two-agent ballistic equilibrium cited in
Contribution 1 of the main text, where strict alternation gives
$\overline{\mathrm{RP}} = 1$ exactly. Departing from the symmetric equilibrium
by letting agent $A$ claim a growing share $w_A > 0.5$ of turns, on an
otherwise maximally regular schedule, drops both standard sub-measures, RS
faster than WPE (at $w_A=0.9$: $\mathrm{RS}=0.111$, $\mathrm{WPE}=0.200$,
$\overline{\mathrm{RP}}=0.156$); rhythm is punished more harshly than frequency
here because agent $A$'s gaps shrink far below the symmetric ideal while agent
$B$'s lengthen far beyond it, a wider symmetric deviation than the
corresponding win-count imbalance. The weighted variants $\mathrm{RS}^w$,
$\mathrm{WPE}^w$, and $\overline{\mathrm{RP}}^w$ ($r_i^* = 1/w_i - 1$),
recalibrated to the claimed share, remain exactly $1$ throughout
(Figure~\ref{fig:exp4}), so RP correctly distinguishes a regular-but-unequal
allocation from an irregular one, and only the uniform-target variant flags the
departure from the symmetric equilibrium.

\section{Reward-Rule Design: Why the Population-Based Rule Is Primary}
\label{sup:reward}

We chose the population-based ILF/IQF rule as primary a priori, before
examining any results, for measurement cleanliness. It keeps the tie reward
deterministic and stationary given only the outcome type (solo win, partial
tie, full collision), so that variation in the metrics is not confounded with
episode-to-episode fluctuation in how many agents happen to collide, a
quantity a per-claimant reward would itself track. It also keeps universal
collision, $k=n$, at exactly zero reward by design, so that total
non-coordination can never register as partial success under outcome-based
metrics such as Efficiency, precisely the gap between traditional and temporal
metrics that this paper's central finding depends on exposing.

As a complementary check using real Q-learning rather than a synthetic
construction, we re-ran the pipeline under the canonical Rosenthal
congestion-game payoff~\cite{rosenthal73}, distinct from the
capacity-threshold structure of market-entry games~\cite{selten82}, where each
of $k$ simultaneous claimants receives an equal share $r_\mathrm{high}/k$, a
smooth function of $k$ alone, in place of this paper's own threshold-based
ILF/IQF rule, which instead fixes a flat share for any partial tie and
collapses only at full occupancy ($k=n$), the step structure more
characteristic of market entry. We ran this check across the full tested
population range, $n \in \{2,3,5,8,10\}$. For $n \geq 3$ the resulting RS, WPE,
$\overline{\mathrm{RP}}$, and CALT all fall within $1$--$2\%$ of the ILF values
reported in the main text, and the gap narrows further, not just remains small,
as $n$ grows (CALT's relative change shrinks monotonically from $-9.8\%$ at
$n=3$ to $-0.2\%$ at $n=8$ and effectively $0\%$ at $n=10$), confirming that
this a priori design choice, sound on its own methodological terms, turns out
not to be empirically load-bearing for the coordination gap either.

The coordination gap is therefore not an artefact of the reward-threshold
choice, and this robustness check reinforces, rather than undercuts, the need
for the ALT/RP framework itself, since it is that framework, not
congestion-game equilibrium theory, that characterises the outcome under either
reward rule. At $n=2$ both CALT and $\overline{\mathrm{RP}}$ drop substantially
under the equal-split rule (CALT: $0.315 \to 0.230$; $\overline{\mathrm{RP}}$:
$0.606 \to 0.325$), a genuine and fully explained difference. With only two
agents, every collision is necessarily an all-agent collision, so the one point
on which the two reward rules disagree, a $0$ payoff under ILF's capacity
collapse versus $r_\mathrm{high}/2$ under the equal split, governs every
collision rather than a minority of them as it does for $n \geq 3$. Both
metrics move in the same direction and shrink toward zero effect with the same
$n$-trend, which is itself a cross-metric consistency check. RP is not merely
rank-correlated with CALT on the single reward regime reported in the main
text, it responds to a genuine reward-structure perturbation the same way CALT
does.

\section{Per-Sub-Metric Correlations and Named RP Variants}
\label{sup:corr}

Table~\ref{tab:rp_variants} names every $\overline{\mathrm{RP}}$ pairing the
main text refers to collectively, and Table~\ref{tab:correlations} gives the
full per-sub-metric correlation matrix that the main text summarises to its
four combined rows.

\begin{table*}[t]
\centering
\caption{Named $\overline{\mathrm{RP}}$ variants. All combine RS and WPE as
  $(\mathrm{RS} + \mathrm{WPE})/2$. Subscripts: excl = exclusive wins,
  reach = all terminal reaches. Spearman $\rho_S$ against CALT / EALT / AALT
  from Table~\ref{tab:correlations} ($N=30$).}
\label{tab:rp_variants}
\small
\begin{tabular}{llll}
\toprule
Variant & RS source & WPE source & $\rho_S$ (CALT / EALT / AALT) \\
\midrule
$\overline{\mathrm{RP}}_\text{excl}$         & excl  & excl  & 0.970 / 0.996 / 0.967 \\
$\overline{\mathrm{RP}}_\text{reach}$        & reach & reach & 0.990 / 0.954 / 0.971 \\
$\overline{\mathrm{RP}}_\text{RS-mxAE}$      & reach & excl  & 0.975 / 0.976 / 0.963 \\
$\overline{\mathrm{RP}}_\text{RS-mxAX}$      & excl  & reach & 0.960 / 0.980 / 0.956 \\
\midrule
FRP (weighted, $w_i=1/n$)  & excl & excl & equal to $\overline{\mathrm{RP}}_\text{excl}$ \\
ERP (weighted, $w_i$ free) & excl & excl & heterogeneous targets \\
\bottomrule
\end{tabular}
\end{table*}

\begin{table}[ht]
\centering
\caption{Spearman $\rho_S$ between every RP-family submetric and the three ALT
  primary metrics ($N = 30$ configurations: 20 Q-learning runs across
  $n \in \{2,3,5,8,10\} \times \{\text{Type-A},\text{Type-B}\} \times
  \{\text{ILF},\text{IQF}\}$, plus 10 random baselines). Superscripts:
  $^*$ $p < 0.001$; $^\dagger$ $p < 0.05$. avg\_wait is the raw mean
  waiting-period length in episodes; its correlation is negative for exclusive
  wins (longer waits mean worse coordination) and positive for all-reaches
  (longer reach-waits mean fewer simultaneous-arrival collisions), a sign flip
  discussed below. AWE$_\text{excl}$ is exactly $0$ for Q-learning at
  $n \geq 3$ but not at $n=2$. All submetrics use the boundary-inclusive
  waiting-period convention and the corrected exactly-one-solo-win AALT
  definition of the main text.}
\label{tab:correlations}
\small
\begin{tabular}{lccc}
\toprule
Metric & CALT & EALT & AALT \\
\midrule
\multicolumn{4}{l}{\textit{Sub-metrics (exclusive wins)}}\\
avg\_wait$_\text{excl}$ & $-0.973^*$ & $-0.936^*$ & $-0.978^*$ \\
AWE$_\text{excl}$ & $0.765^*$ & $0.760^*$ & $0.760^*$ \\
WPE$_\text{excl}$ & $0.970^*$ & $0.996^*$ & $0.967^*$ \\
RS$_\text{excl}$  & $0.959^*$ & $0.986^*$ & $0.950^*$ \\
\multicolumn{4}{l}{\textit{Sub-metrics (all terminal reaches)}}\\
avg\_wait$_\text{reach}$ & $0.508^\dagger$ & $0.594^*$ & $0.471^\dagger$ \\
AWE$_\text{reach}$ & $0.989^*$ & $0.955^*$ & $0.963^*$ \\
WPE$_\text{reach}$ & $0.863^*$ & $0.858^*$ & $0.889^*$ \\
RS$_\text{reach}$  & $0.989^*$ & $0.955^*$ & $0.963^*$ \\
\midrule
\multicolumn{4}{l}{\textit{Combined $\overline{\mathrm{RP}} = (\mathrm{RS} + \mathrm{WPE})/2$}}\\
RS$_\text{excl}$ $+$ WPE$_\text{excl}$   & $0.970^*$ & $0.996^*$ & $0.967^*$ \\
RS$_\text{reach}$ $+$ WPE$_\text{excl}$  & $0.975^*$ & $0.976^*$ & $0.963^*$ \\
RS$_\text{excl}$ $+$ WPE$_\text{reach}$  & $0.960^*$ & $0.980^*$ & $0.956^*$ \\
RS$_\text{reach}$ $+$ WPE$_\text{reach}$ & $0.990^*$ & $0.954^*$ & $0.971^*$ \\
\bottomrule
\end{tabular}
\end{table}

We do not seek a single best submetric here. RS, WPE, AWE and the raw
avg\_wait quantity are all facets of the same lightweight RP family, and the
goal of Table~\ref{tab:correlations} is to show that this \emph{entire family},
under either win-event definition, tracks the expensive ALT family closely
enough to serve as its economical substitute. Several patterns support this.

\textit{Even the rawest, uninterpreted quantity already tracks ALT.} Before any
normalisation into a bounded score, the mean waiting-period length avg\_wait
correlates with all three ALT metrics at $|\rho_S| \geq 0.92$ under the
exclusive-win definition. The sign is negative, as expected, since longer waits
between an agent's sole victories signal worse coordination, hence lower ALT.
Under the all-reaches definition the sign flips to positive
($\rho_S = 0.47$--$0.59$), since a longer average gap between an agent's
terminal reaches (win or tie) means simultaneous-arrival collisions are rarer,
hence \emph{better}, not worse, coordination. This sign reversal is itself
informative. It confirms that exclusive-win and all-reaches events capture two
genuinely different failure modes (individual under-winning versus collective
collision), consistent with the win-event definition discussion in the main
text.

\textit{WPE and RS, under either win-event definition, align strongly with ALT.}
WPE$_\text{excl}$ reaches $\rho_S = 0.970$, $0.996$, $0.967$ (CALT, EALT, AALT)
and RS$_\text{excl}$ reaches $0.959$, $0.986$, $0.950$, both tracking whether
agents receive their fair share of access, via different computational paths
(per-agent frequency and rhythm versus sliding-window batch scoring). Under the
all-reaches definition, RS$_\text{reach}$ and AWE$_\text{reach}$ coincide
exactly ($\rho_S = 0.989$, $0.955$, $0.963$). Reach events are frequent enough
that no agent's average gap ever exceeds AWE's $2r_i^*$ cutoff, so the two
formulas reduce to the same quantity in this regime. AWE$_\text{excl}$, by
contrast, is exactly $0$ for Q-learning at $n \geq 3$; its still-substantial
$\rho_S \approx 0.76$ is driven entirely by variation in the random baselines
and by the $n=2$ Q-learning configurations, where exclusive wins remain
frequent enough to avoid the hard cutoff. That collapse is the reason RS, not
AWE, is the rhythm sub-measure of RP.

\textit{Every combined $\overline{\mathrm{RP}}$ pairing performs comparably well.}
All four excl/reach combinations of RS and WPE achieve $\rho_S \geq 0.95$
against every ALT metric, clustering within a narrow $0.95$--$0.99$ band
regardless of which win-event definition is used for which component. This
robustness is the practical takeaway. A user need not resolve the
exclusive-versus-all-reaches choice precisely to obtain a reliable RP score,
since the family is sufficient to detect deviation from Perfect Alternation
under either convention.

\begin{figure}[ht]
\centering
\includegraphics[width=0.92\linewidth]{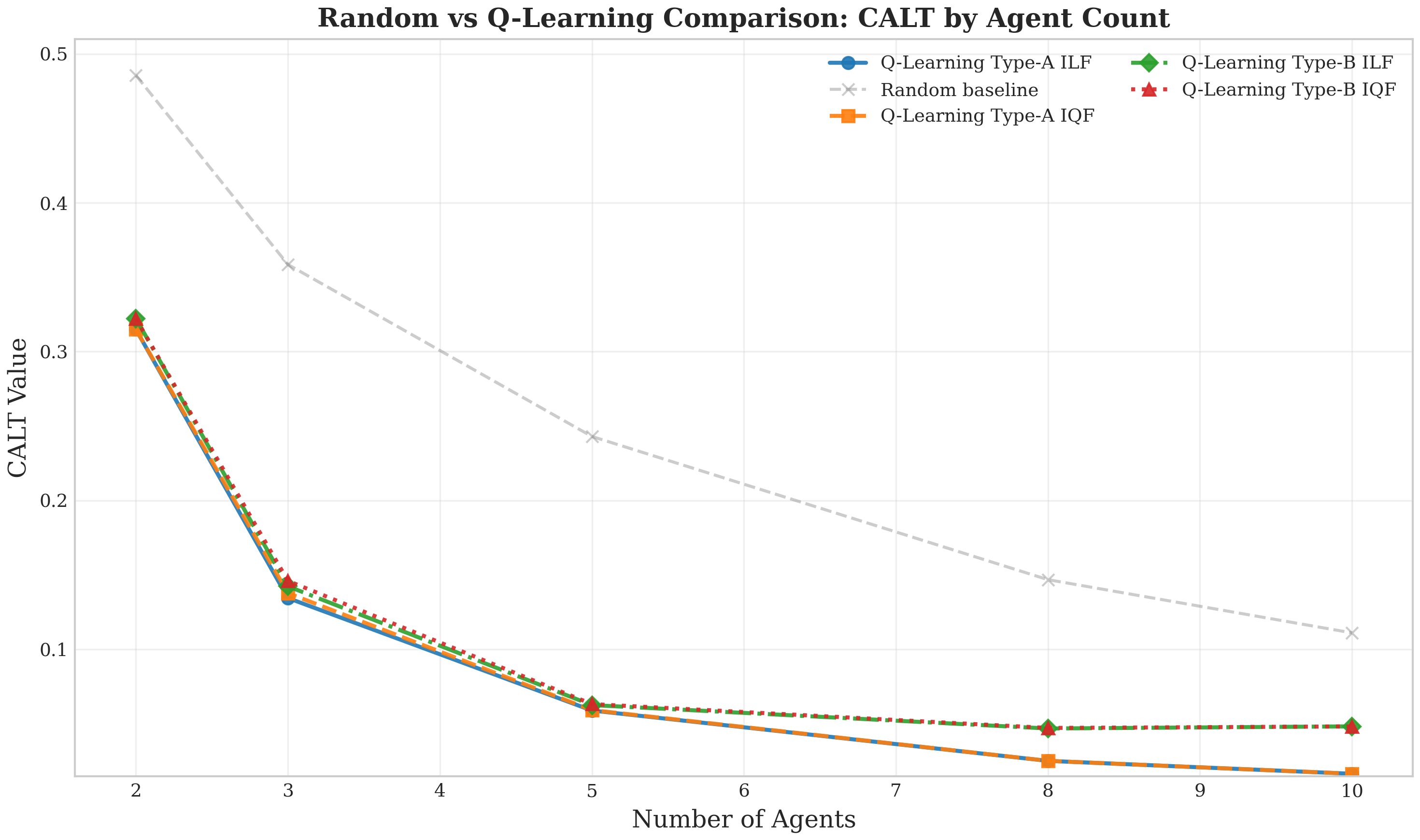}
\caption{CALT values for Q-learning (all four configurations:
  Type-A/B $\times$ ILF/IQF) and the random baseline across agent counts
  $n \in \{2,3,5,8,10\}$. Q-learning agents (solid coloured lines) fall
  consistently \emph{below} the random baseline (grey dashed), revealing
  coordination failure undetectable by traditional metrics.}
\label{fig:metrics}
\end{figure}

Figure~\ref{fig:metrics} shows that the coordination failure reported in the
main text for the Type-A/ILF condition holds across all four Q-learning
configurations against their respective random baselines.

\section{Full Metric Values Across All State and Reward Configurations}
\label{sup:fulltables}

Table~\ref{tab:s1} reports $\overline{\mathrm{RP}}$ (RS$_\text{excl}$ +
WPE$_\text{excl}$), CALT, EALT, and AALT for every Type-A/Type-B $\times$
ILF/IQF Q-learning configuration, plus the random baseline, across all agent
counts $n \in \{2,3,5,8,10\}$. The Type-A/ILF rows reproduce the main text's
primary results table; Type-A/IQF, Type-B/ILF, and Type-B/IQF are reported here
in full. The random baseline is identical under ILF and IQF (a purely random
policy does not condition its action choice on the reward structure), so a
single row per $n$ is given. All values use the boundary-inclusive
waiting-period convention and the corrected exactly-one-solo-win AALT
definition described in the main text.

\begin{table*}[t]
\centering
\caption{Full metric values by agent count $n$, state representation
  (Type-A/Type-B), and reward rule (ILF/IQF). $\nu$ is the episode budget.
  The random baseline uses $\nu=10{,}000$ throughout and does not depend on
  the reward rule.}
\label{tab:s1}
\small
\begin{tabular}{rllrrrrr}
\toprule
$n$ & Policy & State/Reward & $\nu$ & $\overline{\mathrm{RP}}$ & CALT & EALT & AALT \\
\midrule
2  & Q-Learning & Type-A, ILF & 1{,}000   & 0.606 & 0.315 & 0.522 & 0.296 \\
2  & Q-Learning & Type-A, IQF & 1{,}000   & 0.606 & 0.315 & 0.522 & 0.296 \\
2  & Q-Learning & Type-B, ILF & 1{,}000   & 0.618 & 0.322 & 0.513 & 0.284 \\
2  & Q-Learning & Type-B, IQF & 1{,}000   & 0.618 & 0.322 & 0.513 & 0.284 \\
2  & Random     & --          & 10{,}000  & 0.755 & 0.486 & 0.651 & 0.435 \\
\midrule
3  & Q-Learning & Type-A, ILF & 4{,}721   & 0.195 & 0.134 & 0.200 & 0.112 \\
3  & Q-Learning & Type-A, IQF & 4{,}721   & 0.199 & 0.138 & 0.204 & 0.113 \\
3  & Q-Learning & Type-B, ILF & 4{,}721   & 0.198 & 0.142 & 0.203 & 0.115 \\
3  & Q-Learning & Type-B, IQF & 4{,}721   & 0.203 & 0.146 & 0.207 & 0.116 \\
3  & Random     & --          & 10{,}000  & 0.633 & 0.359 & 0.545 & 0.323 \\
\midrule
5  & Q-Learning & Type-A, ILF & 31{,}839  & 0.085 & 0.059 & 0.087 & 0.038 \\
5  & Q-Learning & Type-A, IQF & 31{,}839  & 0.086 & 0.059 & 0.087 & 0.038 \\
5  & Q-Learning & Type-B, ILF & 31{,}839  & 0.098 & 0.062 & 0.099 & 0.043 \\
5  & Q-Learning & Type-B, IQF & 31{,}839  & 0.098 & 0.063 & 0.099 & 0.043 \\
5  & Random     & --          & 10{,}000  & 0.456 & 0.243 & 0.410 & 0.203 \\
\midrule
8  & Q-Learning & Type-A, ILF & 174{,}583 & 0.028 & 0.025 & 0.029 & 0.009 \\
8  & Q-Learning & Type-A, IQF & 174{,}583 & 0.028 & 0.025 & 0.028 & 0.009 \\
8  & Q-Learning & Type-B, ILF & 174{,}583 & 0.107 & 0.047 & 0.109 & 0.035 \\
8  & Q-Learning & Type-B, IQF & 174{,}583 & 0.109 & 0.047 & 0.111 & 0.035 \\
8  & Random     & --          & 10{,}000  & 0.261 & 0.147 & 0.251 & 0.100 \\
\midrule
10 & Q-Learning & Type-A, ILF & 385{,}281 & 0.014 & 0.016 & 0.014 & 0.004 \\
10 & Q-Learning & Type-A, IQF & 385{,}281 & 0.014 & 0.016 & 0.015 & 0.004 \\
10 & Q-Learning & Type-B, ILF & 385{,}281 & 0.162 & 0.048 & 0.166 & 0.042 \\
10 & Q-Learning & Type-B, IQF & 385{,}281 & 0.164 & 0.048 & 0.167 & 0.043 \\
10 & Random     & --          & 10{,}000  & 0.187 & 0.111 & 0.183 & 0.065 \\
\bottomrule
\end{tabular}
\end{table*}

\paragraph{Observations.} The qualitative pattern reported in the main text
for Type-A/ILF holds across all four Q-learning configurations. Every
$\overline{\mathrm{RP}}$, CALT, EALT, and AALT value is substantially below its
corresponding random baseline at every $n$. ILF and IQF are close throughout
(within $1$--$3\%$ of each other at matched $n$ and state type), confirming that
the choice between the two reward rules is not the primary driver of the
coordination failure reported in the main text. Type-A and Type-B agree closely
at $n=2,3$ ($\leq 2\%$ relative difference) but diverge increasingly at larger
$n$. Type-B's $\overline{\mathrm{RP}}$ exceeds Type-A's by about $15\%$ at
$n=5$, a factor of $3.8$ at $n=8$, and a factor of $11.6$ at $n=10$, while
remaining far below the random baseline throughout. This widening gap, rather
than a constant offset, is consistent with the state-space compactness argument
in the Limitations subsection of the main text. Type-B's state space grows with
the extra $n$-bit winner-flag vector on top of Type-A's positions, so at fixed
episode budgets an increasing share of it goes unvisited as $n$ grows,
producing markedly more variable per-agent win patterns than Type-A at the
same $n$.

\section{Reward Shaping and Mechanism Design}
\label{sup:shaping}

The main text reports that Perfect Alternation is not self-enforcing under
independent learning. That leaves open a question this paper does not settle:
whether the same metrics used here to \emph{diagnose} coordination failure
could be used to \emph{prevent} it, by treating them as optimisation
objectives rather than only as measurements.

Incorporating $\overline{\mathrm{RP}}$ into reward shaping would directly
incentivise temporal proportionality, since the metric rewards waiting gaps
close to $n-1$ occurring at the ideal frequency $\nu/n$, which are exactly the
two properties Perfect Alternation requires. Maximising CALT as a shaped
reward should similarly promote exclusive wins over simultaneous arrivals,
since CALT's tie penalty $\sum_k(n-y_k)$ falls as more agents reach the
terminal in the same episode. The systematic study of either connects to the
mechanism design literature on incentivising fair
behaviour~\cite{moulin03}.

Both are conjectures, offered as the natural next step rather than as
results. Nothing in this paper trains an agent against a shaped reward, and
one caution is worth stating in advance, given the paper's own findings. A
metric used as a training signal is no longer an independent measure of the
behaviour it shaped: an agent trained to maximise $\overline{\mathrm{RP}}$
would score well on $\overline{\mathrm{RP}}$ by construction, and would have
to be evaluated on something else, whether an ALT variant it was not trained
on, a held-out construction of the kind used in Section~\ref{sup:synthetic},
or a direct inspection of the realised win sequence. The separation between
the measure and the objective is what makes the coordination gap reported in
the main text meaningful, and it would have to be re-established deliberately
in any shaping study.

\section{Computation Cost: Per-Variant Timings and Scaling}
\label{sup:timings}

Table~\ref{tab:per_variant_times} breaks down the ALT column of the main
text's computation-time table. The six variants are within a factor of $1.2$ of
one another at every agent count, because each one pays for the same sliding-window
extraction over the episode sequence and differs only in the closed-form score
computed per window. CALT is consistently the most expensive and FALT the least.

The last column is the cost of computing all six variants in one pass. It is not
the sum of the six preceding columns, and is much smaller than that sum, because
the shared extraction is performed once and each per-variant timing already
contains it. At $n=10$ the six together cost $16.96$ s against $12.55$ s for the
cheapest one alone. This is reported for completeness rather than as the
quantity of interest, since evaluating a run calls for one ALT score, not six;
the auxiliary variants FALT, qFALT, and qEALT are not this paper's focus.

\begin{table}[ht]
\caption{Per-variant ALT computation time in seconds, single thread, Type-B
  states with the population-based reward, from the same run as the main text's
  table. ``All six'' is the one-pass cost of computing every variant together,
  not the sum of the columns to its left.}
\label{tab:per_variant_times}
\footnotesize
\centering
\setlength{\tabcolsep}{4pt}
\begin{tabular}{crrrrrrr}
\toprule
$n$ & FALT & EALT & qFALT & qEALT & CALT & AALT & All six \\
\midrule
2  & 0.0098 & 0.0100 & 0.0107 & 0.0107 & 0.0137 & 0.0134 & 0.0226 \\
3  & 0.0327 & 0.0332 & 0.0349 & 0.0352 & 0.0443 & 0.0420 & 0.0665 \\
5  & 0.400  & 0.404  & 0.416  & 0.417  & 0.511  & 0.472  & 0.667 \\
8  & 4.29   & 4.31   & 4.37   & 4.38   & 5.15   & 4.73   & 6.05 \\
10 & 12.55  & 12.60  & 12.74  & 12.76  & 14.86  & 13.61  & 16.96 \\
\bottomrule
\end{tabular}
\end{table}

\subsection*{Scaling beyond the tested range}

The five measurements above cannot separate the two quantities that drive cost.
The episode budget is set by a formula in $n$, so $\nu$ and $n$ grow together and
neither can be held still. The ratio of speedup to $n$ falls from $2.95$ at $n=2$ to
$1.96$ at $n=10$, and nothing in that dataset says which variable is responsible.

Timing a metric needs no training, so the two can be separated directly on
synthetic win sequences, at agent counts far past anything a learning run in this
paper reaches. We time one CALT score against one complete $\overline{\mathrm{RP}}$
score, both from the same raw sequence on one machine, using the implementations
of the constructions in Section~\ref{sup:synthetic}. Absolute times below are not
comparable with Table~\ref{tab:computation_times}, which was measured on the original
run's hardware with the production implementation. The two implementations have
different per-operation costs, so they give different constants. Compared at the
same $n=10$, the production implementation gives $1.96$ and these give $0.80$ to
$1.23$. The scaling is what carries across, and the constant is not.

\paragraph{What the operation counts predict.}
ALT performs $\nu - n + 1$ sliding windows of $n$ entries, an operation count of
$(\nu - n + 1)\, n$. RP sweeps the sequence once and does constant work per agent,
giving $\Theta(\nu + n)$. Their ratio is $(\nu - n + 1)\, n / (\nu + n)$, which tends to $n$
once $\nu \gg n$. Two qualifications matter more here than the limit does. The ratio
of running times is that multiplied by the two implementations' per-operation cost
ratio, a constant the complexity does not fix, so what the separation guarantees is
proportionality to $n$ rather than a floor at $n$. And $\nu - n + 1$ is not $\nu$. The
two agree while $\nu$ stays well above $n$, and ALT is undefined altogether for
$\nu < n$, since a window of $n$ episodes does not fit in the sequence.

\paragraph{The episode budget on its own does not matter.}
Holding $n$ at $10$ and varying $\nu$ over a $400$-fold range, from $1{,}000$ to
$400{,}000$, the speedup measures $7.98$, $10.54$, $12.31$, $9.38$, and $10.66$.
Over the upper $80$-fold range the sequence is not monotone in $\nu$, so there is no
drift to attribute to the episode budget. The shortest sequence sits about a fifth
below the rest, repeatably, which is RP's fixed per-call cost weighing more heavily
when there is little sequence to amortise it over, and not a scaling effect. The decline in the main table's
speedup-to-$n$ ratio is therefore an effect of $n$, not of $\nu$, and it reproduces at
fixed $\nu$, where the ratio falls from $2.53$ at $n=2$ to $0.98$ at $n=20$ before
settling. That decline is per-window overhead amortising over larger windows, and
it is a small-$n$ transient rather than a trend.

\paragraph{At a fixed ratio the growth is linear, out to $n=2000$.}
Table~\ref{tab:projection} holds $\nu = 100n$, which keeps $\nu \gg n$ while $n$ grows
$200$-fold. The speedup rises from $8.0\times$ to $1220\times$, and the constant
of proportionality over $n \geq 50$ stays between $0.577$ and $0.657$, a $13\%$ spread.
Linear growth in $n$ is therefore measured over a $40$-fold range in $n$ rather
than inferred from the five coupled points of the main table.

\paragraph{Where it stops, and why that is not a ceiling in the method.}
Fixing $\nu$ instead and pushing $n$ eventually leaves the $\nu \gg n$ regime, and the
speedup then saturates. At $n=2000$ the same sequence length of $50{,}000$ gives
$664\times$ where $200{,}000$ gives $1220\times$. Pressing further makes the point
sharply. At $n=500$, reducing $\nu$ from $500{,}000$ to $600$ collapses the speedup
from $342\times$ to $2.3\times$, because the window count $\nu - n + 1$ falls from
$499{,}501$ to $101$ while RP still pays its per-agent cost. So the condition to
state alongside any large-$n$ figure is a ratio, not an agent count.

\begin{table}[ht]
\caption{Speedup at a fixed $\nu = 100n$, so that $\nu \gg n$ holds throughout. The
  upper block is measured on this machine with the implementations of
  Section~\ref{sup:synthetic}; the lower block extends the measured constant
  $0.61$ and the measured per-operation costs, and is marked because none of it
  is timed. The constant belongs to these implementations, so the projected
  wall-clock times indicate scale rather than a prediction for any particular
  system.}
\label{tab:projection}
\footnotesize
\centering
\setlength{\tabcolsep}{5pt}
\begin{tabular}{crrrrr}
\toprule
$n$ & $\nu$ & ALT time & RP time & Speedup & Speedup$/n$ \\
\midrule
\multicolumn{6}{l}{\emph{Measured}} \\
10    & $10^{3}$   & 0.0026 s & 0.00032 s & $8.0\times$  & 0.798 \\
50    & $5\times10^{3}$ & 0.0489 s & 0.00149 s & $32.8\times$ & 0.657 \\
100   & $10^{4}$   & 0.179 s  & 0.00308 s & $58.0\times$ & 0.580 \\
500   & $5\times10^{4}$ & 5.24 s & 0.0172 s & $305\times$  & 0.609 \\
1000  & $10^{5}$   & 19.6 s   & 0.0340 s  & $577\times$  & 0.577 \\
2000  & $2\times10^{5}$ & 88.2 s & 0.0723 s & $1220\times$ & 0.610 \\
\midrule
\multicolumn{6}{l}{\emph{Projected, not timed}} \\
$10^{4}$ & $10^{6}$ & 37 min & 0.4 s  & $6.1\times10^{3}$ & 0.61 \\
$10^{5}$ & $10^{7}$ & 2.6 days & 3.6 s & $6.1\times10^{4}$ & 0.61 \\
$10^{6}$ & $10^{8}$ & 255 days & 36 s  & $6.1\times10^{5}$ & 0.61 \\
\bottomrule
\end{tabular}
\end{table}

At the largest projected size the separation stops being a matter of efficiency.
One ALT score over a sequence of $10^{8}$ episodes among $10^{6}$ agents is a
months-long single-threaded computation, while one RP score over the same sequence
is well under a minute. At that size the precise family stops being available at
all, which is the case for a proxy that keeps its ranking.
A vectorised ALT would lower the constant and a vectorised RP would raise it. The
linear growth in $n$ survives either, which is why the constant is reported
separately from the law it multiplies.

\end{document}